\documentclass{osa-article}

\usepackage{cite}
\usepackage{amsmath, amsfonts, amssymb}
\usepackage{graphicx}
\usepackage[caption=false]{subfig}
\usepackage{algorithm,algorithmic}
\usepackage{hyperref}
\usepackage{array}
\usepackage{siunitx}
\usepackage{color}
\usepackage{url}
\usepackage{lipsum, color}
\usepackage{mathtools}

\definecolor{blue}{rgb}{0,0,0.7}
\definecolor{magenta}{cmyk}{0,1,0,0}
\definecolor{darkgreen}{rgb}{0,0.6,0}
\definecolor{purple}{cmyk}{0.5,1,0,0}

\articletype{Research Article}

\begin{document}

\pagestyle{plain}

\title{A Physical Light Transport Model for Non-Line-of-Sight Imaging Applications}

\author{Syed Azer Reza,\authormark{1} Marco La Manna,\authormark{1}, and Andreas Velten\authormark{1,2,*}}

\address{\authormark{1} Department of Biostatistics and Medical Informatics, University of Wisconsin-Madison, 1300 University Avenue, Madison, WI, 53706, USA\\
\authormark{2}Department of Electrical Engineering, University of Wisconsin-Madison, Madison, WI, 53706, USA}

\email{\authormark{*}velten@wisc.edu} 



\begin{abstract}
Non-line-of-sight (NLOS) imaging has recently attracted a lot of interest from the scientific community. The goal of this paper is to provide the basis for a comprehensive mathematical framework for NLOS imaging that is directly derived from physical concepts. We introduce the irradiance phasor field ($\mathcal{P}$-field) as an abstract quantity for irradiance fluctuations, akin to the complex envelope of the Electrical field (E-field) that is used to describe propagation of electromagnetic energy. We demonstrate that the $\mathcal{P}$-field propagator is analogous to the Huygens-Fresnel propagator that describes the propagation of other waves and show that NLOS light transport can be described with the processing methods that are available for LOS imaging. We perform simulations to demonstrate the accuracy and validity of the $\mathcal{P}$-field formulation and provide experimental results to demonstrate a Huygens-like $\mathcal{P}$-field summation behavior.
\end{abstract}

\section{Introduction}
\label{sec:intro}
\subsection{Background}

\begin{figure}[!t]
  \centering
  \subfloat[]{
  \label{fig:los}
  \includegraphics[width=0.45\linewidth]{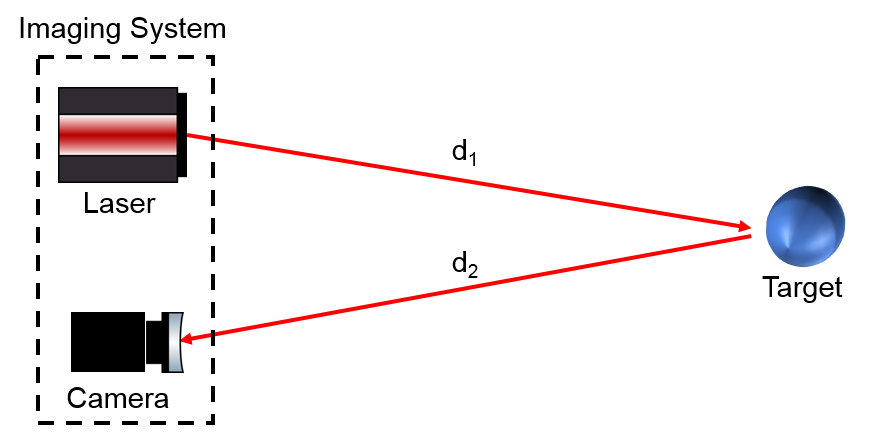}}
  \hfill
  \subfloat[]{
  \label{fig:nlos}
  \includegraphics[width=0.45\linewidth]{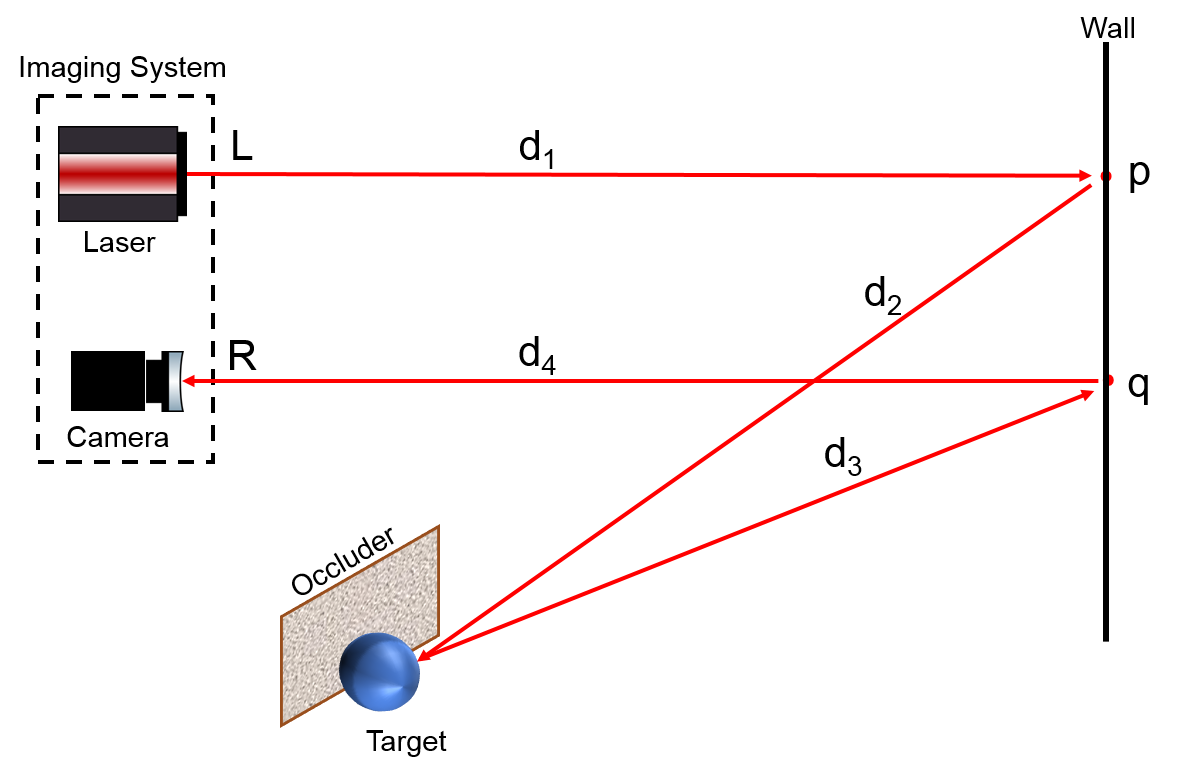}}
  \caption{A LOS scenario, shown in (a), is characterized by a target (a sphere) that is in direct path of both the transmitter (laser) and the receiver (camera). In a NLOS scenario, shown in (b), the target is hidden by an occluder from both the transmitter and the receiver.}
  \label{fig:los_nlos}
\end{figure}

In an optical line-of-sight (LOS) imaging scenario, such as the one depicted in Fig.~\ref{fig:los}, the goal is to reconstruct an image of a target that is in the direct path of a laser source and a camera. Conversely, in an optical non-line-of-sight (NLOS) imaging scenario, the goal is to reconstruct an image of a target that is not in the direct path of the transmitter and receiver. Considering the scene shown in Fig.~\ref{fig:nlos} it has been shown \cite{Velten12,Buttafava15,ramesh20085d, katz2012, kadambi16, balaji2018} that 3D NLOS image recovery can be achieved by illuminating a relay surface in the visible scene (i.e. a relay wall) and collecting light reflected from the object via the relay surface. In Fig.~\ref{fig:nlos}, a light pulse generated by the laser L, incident on a relay wall at point $p$, subsequently scatters from the wall in all directions with a fraction of the photons reflected from the relay wall reaching the target. A fraction of the photons reflected from the target travels back to the wall. The ultra-fast camera, focused at location $q$ on the relay wall, measures the photon flux from the target reflected at $q$ as a function of time.

Reconstructing a 3D image of the hidden object is an inverse light transport problem. Different light transport models describe the propagation of light through a scene (for example, in \cite{Sen05}) and enable us to infer about the scene by analyzing the data captured by a camera~\cite{Seitz05}. Previous approaches have used ray optics and attempted to model light propagation through a scene as a linear operator that can be inverted with a variety of inverse methods~\cite{Kirmani11,Velten12,Gupta12,Wu12, Heide13,Heide14,Heide15,Laurenzis14,Buttafava15,Tsai17,LaManna17}. Non-linear inverse methods for more complex scenes have also been proposed \cite{kadambi16,iseringhausen_18,heide_17}, but the added level of complexity makes their application challenging. Questions regarding null-spaces, attainable resolutions and contrast and dealing with multiple reflections in the hidden scene have also been discussed in prior art. For example, the role of the BRDF and effects of null-spaces have been discussed in \cite{heide_17} and \cite{liu_19CVPR} respectively. Model complexity and inaccurate modeling of real light transport pose a great challenge for conducting more fundamental analyses on NLOS imaging. 

In some of the recent works \cite{liu2018virtual, lindell2019wave}, a wave propagation-based $\mathcal{P}$-field approach for NLOS imaging has produced excellent image reconstruction results of hidden scenes. Experimental results in \cite{liu2018virtual} demonstrate the benefits in treating an NLOS imaging system as a virtual camera-based $\mathcal{P}$-field imaging system. The goal of this paper is to describe ToF NLOS light transport using a wave propagation model akin to those governing other imaging methods. Here, we describe the fundamental mathematical foundations of $\mathcal{P}$-field summation for imaging applications, the limitations of the $\mathcal{P}$-field NLOS imaging approach and the requisite assumptions and approximations for the validity of the $\mathcal{P}$-field model. 

The newly introduced $\mathcal{P}$-field approach denotes the complex envelope of the average optical irradiance. In this paper we show that propagation of $\mathcal{P}$-fields from the virtual aperture at the relay wall to a virtual sensor behind the wall can be modeled --analogous to the Huygens' integral-- as the propagation of wave-like $\mathcal{P}$-field wavelets contributions from the aperture plane to the detector plane. With the proposed \emph{phasor field} ($\mathcal{P}$-field) (A phasor representation of radiance was discussed in~\cite{MGupta15}, where the authors propose a framework to analyze the light transport in correlation-based TOF ranging) formalism, we show that NLOS imaging can be treated similarly to LOS imaging.  

With the aid of simulation results, we also demonstrate the effect of aperture roughness (here roughness refers the material roughness of the relay wall, which we treat here as the $\mathcal{P}$-field virtual aperture) on the accuracy of the amplitude and phase estimates in the detector plane -- obtained through the $\mathcal{P}$-field integral which involves a summation of $\mathcal{P}$-field contributions from the aperture. We also present preliminary experimental results where we implemented a $\mathcal{P}$-field interferometer and measured the change in $\mathcal{P}$-field signal amplitude by changing the path length of one interferometer arm while keeping the path length of the other reference arm fixed.

\section{The $\mathcal{P}$-field Imaging Approach}
\label{sec:P_Field_App}
\subsection{The $\mathcal{P}$-field Integral}
\label{sec:P_Field_Int}

\begin{figure*}
  \centering
  \subfloat[]{
  \label{fig:huygens_a}
  \includegraphics[width=0.45\linewidth]{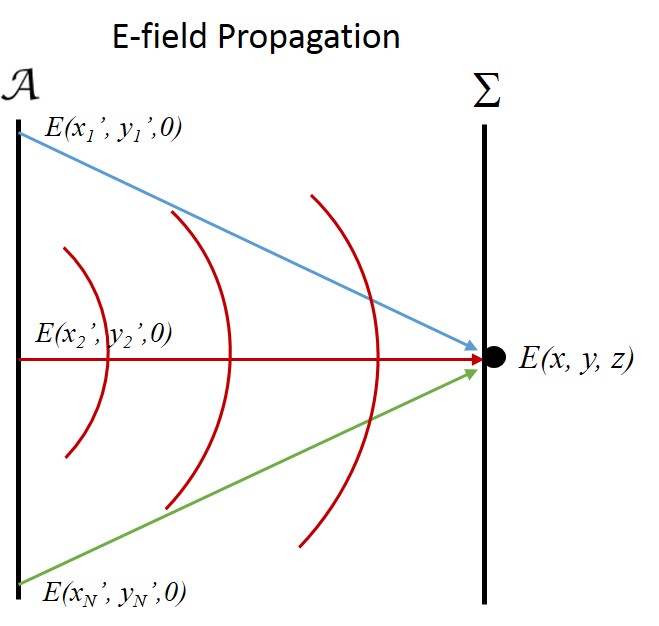}} 
  \hfill  
  \subfloat[]{
  \label{fig:huygens_b}
  \includegraphics[width=0.45\linewidth]{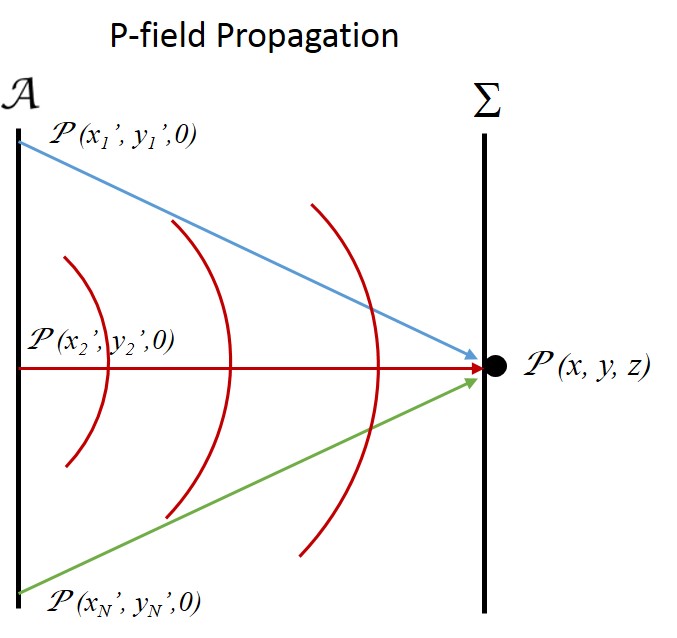}}
  \caption{Analogy between E-field and $\mathcal{P}$-field propagation: Scalar E-field contributions $E(x',y',0)$ (top) and $\mathcal{P}$-field contributions (bottom) from all generic locations $(x',y',0)$ on an aperture plane $\mathcal{A}$ to a location $(x,y,z)$ on an observation plane $\Sigma$ where $\mathcal{A}$ and $\Sigma$ are separated by a distance $z$ and any pair of locations $(x',y',0)\in \mathcal{A}$ and $(x,y,z)\in \Sigma$ are separated by a distance $|r|$.} 
  \label{fig:huygens}
\end{figure*}

To explain what we mean by a Huygens'-like integral describing the propagation of $\mathcal{P}$-fields, let us consider Fig.~\ref{fig:huygens} which describes the propagation of various E-field spherical wavelet contributions from an aperture plane $\mathcal{A}$ to a detection plane $\Sigma$ separated by an arbitrary distance $z$. The Green's function-based solution of the wave equation describes the propagation of scalar E-field wavelet contributions from each location $(x',y',0) \in \mathcal{A}$ to any particular location $(x, y, z) \in \Sigma$ with the resulting scalar E-field $E(x,y,z)$ at $(x, y, z)$ described as a linear sum of these E-field wavelet contributions. The Huygens' integral is given by~\cite{Goodman96,Stutzman98}
\begin{equation}
 \label{eq:E_wav3}
 E(x,y,z)= K_\mathrm{E} \iint\limits_{\mathcal{A}} \tau (x', y',0) \, E_0(x',y',0) \frac{e^{j K |r|}}{|r|} \, \chi \, dx' dy'.
\end{equation}
In \eqref{eq:E_wav3}, $K_\mathrm{E}=1/j\lambda_\mathrm{E}$ is the E-field proportionality coefficient, $\tau(x',y',0) = t_0(x',y',0)e^{j\Delta\phi_{K}(x',y',0)}$ and $E_0(x', y',0)$ are the E-field transmission function and the amplitude of the E-field contribution from location $(x',y',0) \in \mathcal{A}$ respectively, $K = 2 \pi / \lambda_\mathrm{E}$ is the E-field wavenumber, $\lambda_\mathrm{E}$ is the E-field wavelength, $|r|$ is the Euclidian distance between a single location $(x',y',0)\in\mathcal{A}$, $(x, y, z)\in\Sigma$ expressed as $ |r| = \sqrt{(x-x')^2 + (y-y')^2 + z^2}$, and $\chi$ is the obliquity factor; hereon we assume that $\chi = 1$. Moreover, in \eqref{eq:E_wav3},
\begin{equation}
 \label{eq:E_wav2}
 \mathcal{E} = t (x', y',0) E_0 (x', y',0) \frac{e^{j K |r|}}{|r|},
\end{equation}
is defined as an E-field spherical wavelet contribution from $(x',y',0)\in \mathcal{A}$. 

In our case, the optical irradiance is modulated by a time harmonic function $P(t)$ (which we refer to as the $\mathcal{P}$-field signal) with frequency $\Omega$ and a corresponding wavelength $\lambda_P$. Let us assume that the following condition on the coherence length $l_\mathrm{c}$ of light holds true 
\begin{equation}
    \label{eq:Coh_length}
    l_\mathrm{c} \ll \lambda_\mathrm{P}.
\end{equation}
Moreover, let 
\begin{equation}
    \label{eq:int_time2}
    \frac{2\pi}{\omega} \ll T \ll \frac{2\pi}{\Omega}, 
\end{equation}
where the detector integration time is much longer than the time period of the short optical carrier (E-field) of frequency $\omega$ but much shorter than the longer time period of the $\mathcal{P}$-field signal of frequency $\Omega$. We show in the Appendix that we can describe $\mathcal{P}$-field propagation and summation in $\Sigma$ as the sum of Huygens'-like $\mathcal{P}$-field wavelet contributions from $\mathcal{A}$. Namely, similarly to \eqref{eq:E_wav3} for E-fields, the temporal evolution of the time-average irradiance can be described by the $\mathcal{P}$-field integral as a sum of $\mathcal{P}$-field wavelet contributions $\mathcal{P}(r)$ from $\mathcal{A}$ as 
\begin{equation}
 \label{eq:PhasorWav}
 |I_\mathrm{Tot-F} (x,y,z)| = K_\mathrm{P} \bigg| \iint\limits_{\mathcal{A}} \frac{1}{|r|} \underbrace{\mathcal{P}_{0,\Omega}(x',y',0) \frac{e^{j \beta |r|}}{|r|}}_{\mathcal{P}(r)} \chi \, dx'dy' \bigg|,
\end{equation}
where $\mathcal{P}(r)$ represents the phasor form of a signed, real, time harmonic, monochromatic $\mathcal{P}$-field contribution $\mathcal{P}(r,t)$ stated as $\mathcal{P}(r,t)=\mathrm{Re}[\mathcal{P}(r)e^{j\Omega t}]$. In \eqref{eq:PhasorWav}, $|I_\mathrm{Tot-F} (x,y,z)|$ is the sum of the signed envelopes of optical irradiance at $(x,y,z)$ described as the sum of $\mathcal{P}$-field wavelet contributions from all locations $(x',y',0)\in \mathcal{A}$ with respective amplitudes of $\mathcal{P}_{0,\Omega}(x',y',0)$, $K_\mathrm{P}$ is the $\mathcal{P}$-field proportionality coefficient, $\beta$ is the $\mathcal{P}$-field wavenumber expressed either in terms of the $\mathcal{P}$-field wavelength $\lambda_\mathrm{P}$ or its frequency $\Omega$ (where $n$ denotes the refractive index of the medium between $\mathcal{A}$ and $\Sigma$, and $c$ denotes the speed of light in vacuum) as 
\begin{equation}
 \label{eq:beta}
 \beta = \frac{2\pi}{\lambda_\mathrm{P}} = \frac{n \Omega}{c}.
\end{equation}

In other words $\mathcal{P}(r,t)$ describes the irradiance at a location $(x,y)$ as a function of time $t$ and distance $|r|$ with the static background irradiance component subtracted allowing $\mathcal{P}(r,t)$ to be a signed quantity. Just like the E-field, the $\mathcal{P}$-field $\mathcal{P}(x',y',z',t)$ at any location $(x',y',z')$ can be expressed as a superposition of monochromatic phasor components:
\begin{equation}
    \mathcal{P}(x',y',z',t)=\int_{-\infty}^{\infty}{\mathcal{P}_{\Omega}(x',y',z',t)d\Omega}
\end{equation}
where $\mathcal{P}_\Omega(x',y',z',t) = \mathcal{P}_{0,\Omega}(x',y',z')e^{-i\Omega t}$ is a monchromatic $\mathcal{P}$-field wave and $\mathcal{P}_{0,\Omega}(x',y',z') = \int_{-\infty}^{\infty} \mathcal{P}(x',y',z',t)e^{j\Omega t} dt$ is the Fourier domain representation of $\mathcal{P}(x',y',z',t)$. 

Analogous to the complex E-field, $P_\Omega(x',y',z',t)$ is symmetric about the origin $\Omega=0$ resulting in $P(x',y',z',t)$ to be always real. In practice we usually omit the negative frequency component and compute only the complex phasor, keeping in mind that the actual phasor field intensity is given by the sum of the positive and negative frequency components (i.e. the phasor and its complex conjugate). Note also that other than the electric field, the intensity cannot be negative which only affects the phasor at $P_{\Omega=0}$. This static component of the field can usually safely be ignored. We can also subtract it and consider only the fast variations: $P'_{\Omega}(x',y',z',t) = P_{\Omega}(x',y',z',t) - P_{\Omega=0}(x',y',z',t)$. In the following we consider a single monochromatic component $P_\Omega(x',y',z',t)$ and this distinction is of no consequence.

To improve readability and maintain clarity throughout the manuscript, a complete derivation of the $\mathcal{P}$-field integral is presented in Appendix -- where we provide details of all requisite approximations and assumptions about aperture roughness, coherence length and detector integration time, which allow us to arrive at the result in \eqref{eq:PhasorWav}. Notations used for describing quantities in the E-field Huygens' integral and the Huygens'-like $\mathcal{P}$-field integral are summarized in Table~\ref{tab:notations1}. Moreover, to avoid confusion, all symbols used throughout the manuscript are summarized in Table~\ref{tab:notations2}.
\begin{table}[!ht]
 \renewcommand{\arraystretch}{1.2}
 \caption{Comparison of E-field and $\mathcal{P}$-field notations.}
 \centering
 \begin{tabular}{ |p{0.17\linewidth}|p{0.3\linewidth}|p{0.4\linewidth}| }  
 \hline
 \hline
  & Carrier electric field & phasor field \\
 \hline
 \hline
  Wavelet & $\mathcal{E}(r) = E_0(x',y',0)\frac{e^{jK|r|}}{|r|}$ & $\mathcal{P} (r) = \mathcal{P}_{0,\Omega}(x',y',0)\frac{e^{j\beta|r|}}{|r|}$ \\
  & & $\mathcal{P}_\Omega(x',y',t) = \mathcal{P}_{0,\Omega}(x',y')e^{-i\Omega t}$ \\
  & & $\mathcal{P}_{0,\Omega}(x',y') = \int_{-\infty}^{\infty} \mathcal{P}(x',y',t)e^{j\Omega t} dt$ \\
  \hline
 Peak Amplitude & $E_0$ & $\mathcal{P}_{0,\Omega}$\\ 
  \hline
 Proportionality Coeff. & $K_\mathrm{E}$ & $K_\mathrm{P}$ \\ 
  \hline
 Wavelength (m) & $\lambda_\mathrm{E}$ & $\lambda_\mathrm{P}$ \\
  \hline
  Frequency (Hz) & $\omega$ & $\Omega$ \\
  \hline
  Wavenumber ($m^{-1}$) & $K=2\pi/\lambda_\mathrm{E}=\omega/c$ & $\beta = 2\pi/\lambda_\mathrm{P}=\Omega/c$ \\
  \hline
 \end{tabular}
 \label{tab:notations1}
\end{table}
\begin{table}[!ht]
 \caption{Summary of coefficients and symbols.}
 \renewcommand{\arraystretch}{1.5}
 \centering
 \begin{tabular}{ |c|c| }  
  \hline
  Symbol & Description \\
  \hline
  \hline
  $\gamma$ (m)  & Maximum aperture roughness \\
  \hline
  $\mathcal{L}$ & Irradiance loss/scattering coefficient   \\
  \hline
  $C(x,y,z,|r|)$ & Ultra near-field correction factor \\
  \hline
  $\langle \eta \rangle$ & Estimation error for $\gamma \approx 0$   \\
  \hline
  $\zeta$ & Medium impedance  \\
  \hline
 \end{tabular}
\label{tab:notations2}
\end{table}
%

\subsection{Correcting Amplitude Error of $\mathcal{P}$-field Integral}
\label{sec:Corr_fac}

Note that the spherical $\mathcal{P}$-field wavelet contributions in \eqref{eq:PhasorWav} are not the only term in the $\mathcal{P}$-field integral argument as an additional $1/|r|$ term multiplies with each corresponding $\mathcal{P}$-field wavelet contribution term. For cases when either of the respective far-field (Fraunhofer) or the near-field (Fresnel) approximations in \eqref{eq:Fraun_cond_a} and \eqref{eq:Fres_cond_a} hold,
(for a chosen $\mathcal{P}$-field wavelength $\lambda_\mathrm{P}$), we can assume that $1/|r|\approx 1/z$. For these cases, \eqref{eq:PhasorWav} can be expressed as
\begin{equation}
    \label{eq:time_5}
    |I_\mathrm{Tot-F}(x,y,z)| = \bigg|\frac{K_\mathrm{P}\mathcal{L}}{z} \int_{\mathcal{A}} \mathcal{P}_{0,\Omega}(x',y',0)\frac{e^{j\beta |r|}}{|r|} \chi dx'dy'\bigg|.
\end{equation}
In the case of ultra near-field imaging scenario where neither of the conditions in \eqref{eq:Fraun_cond_a} and \eqref{eq:Fres_cond_a} are satisfied, $1/|r|\not\approx 1/z$. In this case simple summation of $\mathcal{P}$-field wavelets would lead to an amplitude estimation error (note there is no phase estimation error even for the ultra near-field case as no approximations are made to the phase terms). For the ultra near-field case, the introduction of an amplitude correction factor $C(x,y,z,|r|)$ allows us to correct for the aforementioned amplitude estimation error and enables us to accurately express \eqref{eq:time_5} as a sum of $\mathcal{P}$-field wavelets
\begin{equation}
    \label{eq:time_6}
    |I_\mathrm{Tot-F}(x,y,z)| = \bigg|\frac{K_\mathrm{P}\mathcal{L}C(x,y,z,|r|)}{|r(x,y,z)|_{\mathrm{Av}}} \int_{\mathcal{A}} \mathcal{P}_{0,\Omega}(x',y',0)\frac{e^{j\beta |r|}}{|r|} \chi dx'dy'\bigg|.
\end{equation}
In \eqref{eq:time_6}, the distance
\begin{equation}
    \label{eq:av_dist}
    |r(x,y,z)|_{\mathrm{Av}} = \sqrt{[z^2 + (x -\langle x'\rangle)^2 + (y -\langle y'\rangle)^2]}
\end{equation}
is the distance of any given location $(x,y,z) \in \Sigma$ from an average location $(\langle x'\rangle, \langle y'\rangle , 0)$ in $\mathcal{A}$. In short, this location-dependent correction factor allows us to equate
\begin{equation}
    \label{eq:Corr_fac1}
     \int_{\mathcal{A}} \frac{1}{|r|}\mathcal{P}_{0,\Omega}(x',y',0)\frac{e^{j\beta |r|}}{|r|}\chi dx'dy' 
    = \frac{C(x,y,z,|r|)}{|r(x,y,z)|_{\mathrm{Av}}} \int_{\mathcal{A}}  \mathcal{P}_{0,\Omega}(x',y',0)\frac{e^{j\beta |r|}}{|r|} \chi dx'dy' 
\end{equation}
which enables expressing the $\mathcal{P}$-field integral as pure summation of $\mathcal{P}$-field spherical contributions even for the ultra near-field scenario. This formulation allows to describe $\mathcal{P}$-field propagation from an initial aperture plane to a final imaging plane analogously to how the Huygens' E-field integral completely describes E-field propagation. 

It also has to be noted that we defined the $\mathcal{P}$-field as a single frequency (monotonic) function but it was only done to simplify our mathematical treatment. The $\mathcal{P}$-field imaging integral is correct and applicable even if $\mathcal{P}$-fields represent more complex irradiance fluctuations. Through a spectral decomposition such as a Fourier series representation, any $\mathcal{P}$-field signal with a multi-frequency spectral composition can be simply expressed as a linear summation of single frequency contributions. 

We can also calculate the approximate mean percentage error $\langle\eta\rangle$ (for $m\times n$ total ($x$,$y$) locations in $\Sigma$) between the true amplitude corrected estimate of $|I_\mathrm{Tot}(x,y,z)|$ from \eqref{eq:PhasorWav} and the corresponding uncorrected $\mathcal{P}$-field estimate from \eqref{eq:time_5} (assuming no aperture roughness) for any separation distance $z$ between $\mathcal{A}$ and $\Sigma$. $\langle\eta\rangle$ is given by
\begin{equation}
\label{eq:eta1}
\langle\eta\rangle = \left[ \frac{1}{m} \frac{1}{n} \sum_{u=1}^{m} \sum_{v=1}^{n} \frac{\left| \iint\limits_{\mathcal{A}} \frac{1}{|r_{u,v}|} P_{0,\Omega}\frac{e^{j\beta|r_{u,v}|}}{|r_{u,v}|} dx'dy'\right|- \frac{1}{z} \left|  \iint\limits_{\mathcal{A}}P_{0,\Omega}\frac{e^{j\beta|r_{u,v}|}}{|r_{u,v}|} dx'dy'\right|}{\left| \iint\limits_{\mathcal{A}} \frac{1}{|r_{u,v}|} P_{0,\Omega}\frac{e^{j\beta|r_{u,v}|}}{|r_{u,v}|} dx'dy'\right|}\right] \times 100,
\end{equation}
where the subscripts $u$ and $v$ denote the $u^{th}$ $x$-location and $v^{th}$ $y$-location $\in \Sigma$ and
\begin{equation}
\label{eq:rdef2}
|r_{u,v}|=\sqrt{[z^2 + (x_u -x')^2 + (y_v -y')^2]}.
\end{equation}
The definition of $\langle\eta\rangle$ can be modified to replace the $1/z$ term in \eqref{eq:eta1} by $1/|r(x,y,z)|_\mathrm{AV}$ -- where $|r(x,y,z)|_\mathrm{AV}$ was introduced in \eqref{eq:av_dist} and denotes the distance of any generic location $(x,y,z)\in\Sigma$ from an average location $(\langle x' \rangle, \langle y' \rangle, 0) \in \mathcal{A}$. If the average $\mathcal{P}$-field amplitude estimation error is defined this way, $\langle\eta\rangle = \langle\eta_\mathrm{AV}\rangle$ expressed as
\begin{equation}
\label{eq:eta2}
\langle\eta_\mathrm{AV}\rangle = \left[ \frac{1}{m} \frac{1}{n} \sum_{u=1}^{m} \sum_{v=1}^{n} \frac{\left| \iint\limits_{\mathcal{A}} \frac{1}{|r_{u,v}|} P_{0,\Omega}\frac{e^{j\beta|r_{u,v}|}}{|r_{u,v}|} dx'dy'\right|- \frac{1}{|r(x,y,z)|_{\mathrm{Av}}} \left|  \iint\limits_{\mathcal{A}}P_{0,\Omega}\frac{e^{j\beta|r_{u,v}|}}{|r_{u,v}|} dx'dy'\right|}{\left| \iint\limits_{\mathcal{A}} \frac{1}{|r_{u,v}|} P_{0,\Omega}\frac{e^{j\beta|r_{u,v}|}}{|r_{u,v}|} dx'dy'\right|}\right] \times 100.
\end{equation}
%

\section{Phasor Field Simulations}
\label{sec:simulations}

In this section, we present simulation results of the expected $\mathcal{P}$-field sums at all locations in $\Sigma$. The surface roughness is assumed to be uniformly distributed between 0 and $\gamma$. The computed $\mathcal{P}$-field sum for a monochromatic optical carrier modulated by a monotonic RF signal is normalized to the maximum value of $|I_\mathrm{Tot-F}(x,y,z)|$ (denoted by the subscript 'Norm') which assumes a maximum surface roughness $\gamma \approx 0$. As the $\mathcal{P}$-field integral in \eqref{eq:time3_2} assumes negligible surface roughness, the accuracy of the $\mathcal{P}$-field estimate is affected for apertures with larger roughness - specially if the roughness is comparable to the $\mathcal{P}$-field wavelength. 

The first two simulations (namely Simulation 1 and Simulation 2), performed for the far-field and near-field imaging scenarios, solely investigate this effect on $\mathcal{P}$-field distribution estimate with increasing aperture roughness. In the third and final simulation (Simulation 3), we only investigate the effect of amplitude estimation error for the ultra near-field scenario. In this simulation, we assume negligible aperture roughness (i.e., $\gamma \approx 0$) such that the roughness is enough to scatter light but not enough to cause a significant phase shift to the $\mathcal{P}$-field (which is the underlying assumption for the $\mathcal{P}$-field integral).    

For the first two simulation scenarios, we show that estimates of $\mathcal{P}$-field distributions calculated through the $\mathcal{P}$-field integral in \eqref{eq:time3_2} remain accurate even for high aperture roughness values and gradually degrade as $\gamma \rightarrow \lambda_P$. For Simulation 3, we compare the $\mathcal{P}$-field distribution estimate obtained from the $\mathcal{P}$-field integral in \eqref{eq:time_3_3} without amplitude correction and the $\mathcal{P}$-field integral in \eqref{eq:time_7} with the amplitude correction factor applied. We also show that the $\mathcal{P}$-field estimation error in the absence of significant aperture roughness for the ultra near-field case is only an amplitude error and not a phase estimation error. For each of the three simulations, we set the E-field and $\mathcal{P}$-field wavelengths to $\lambda_\mathrm{E} = \SI{1}{\mu m}$ and $\lambda_\mathrm{P} \approx \SI{30}{cm}$ (corresponding to $\Omega = \SI{1}{GHz}$). Also, for each simulation, we assume a uniform illuminated aperture with arbitrarily chosen dimensions of $\SI{8}{m}\times \SI{4}{m}$.

\subsection{Simulation 1: Far-Field Imaging Scenario}
\label{sec:sim1}

%
\begin{figure}[pt]
  \centering
  \subfloat[]{
  \label{fig:sim1}
  \includegraphics[width=0.45\linewidth]{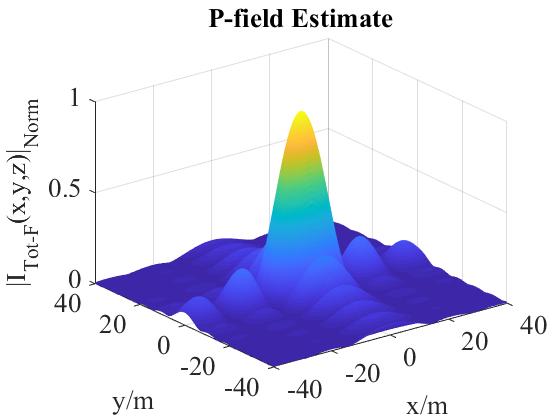}}
  \hfill
  \subfloat[]{
  \label{fig:sim4}
  \includegraphics[width=0.45\linewidth]{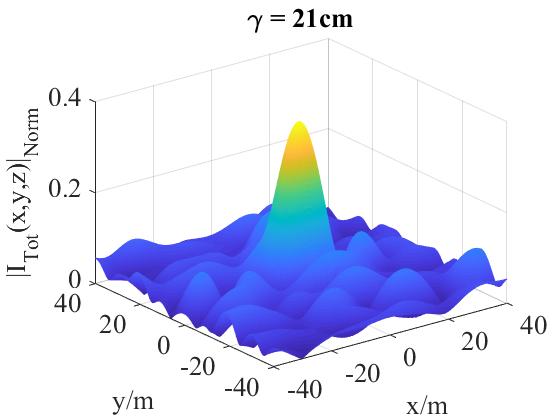}}
  \hfill
  \caption{Far-field results for Simulation 1: (a) $\mathcal{P}$-field estimate $|I_{\mathrm{Tot-F}}(x,y,z)|_{\mathrm{Norm}}$ at $z = \SI{200}{m}$ from the $\mathcal{P}$-field integral of \eqref{eq:time3_2}, and (b) estimate $|I_{\mathrm{Tot}}(x,y,z)|_{\mathrm{Norm}}$ from \eqref{eq:time_7} with roughness of $\gamma = \SI{12}{cm}$ taken into account.}
  \label{fig:simulations}
    \includegraphics[width=0.5\linewidth]{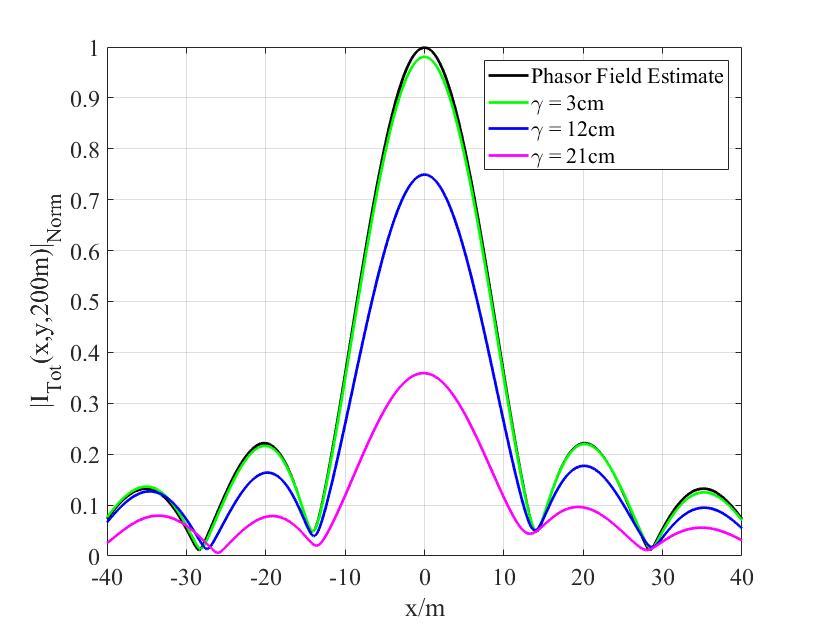}
  \caption{$\mathcal{P}$-field distribution estimate $|I_\mathrm{Tot-F}(x,y,z)|_{\mathrm{Norm}}$ in $\Sigma$ from \eqref{eq:time3_2} and its deviation from actual $\mathcal{P}$-field distribution estimates $|I_{\mathrm{Tot}}(x,y,z)|_{\mathrm{Norm}}$ from \eqref{eq:time_7} at $(x,0,\SI{200}{m})$ with increasing aperture roughness. Plots shown are for surface roughness values of $\gamma = \SI{3}{cm}$, $\gamma = \SI{6}{cm}$, $\gamma = \SI{9}{cm}$, and $\gamma = \SI{15}{cm}$ for a $\mathcal{P}$-field wavelength of $\lambda_\mathrm{P} = \SI{30}{cm}$.}
  \label{fig:varym}
\end{figure}

For this simulation we chose $z = \SI{200}{m}$ as the mutual separation between $\Sigma$ and $\mathcal{A}$. The $\mathcal{P}$-field distribution $|I_\mathrm{Tot-F}(x,y,z)|_\mathrm{Norm}$ estimated from the $\mathcal{P}$-field integral in \eqref{eq:time3_2} is plotted in Fig.~\ref{fig:sim1}. This $\mathcal{P}$-field estimate assumes negligible roughness and hence does not depend on any changes to the aperture roughness. We compare this $\mathcal{P}$-field estimate to the actual $\mathcal{P}$-field distribution $|I_\mathrm{Tot}(x,y,z)|_\mathrm{,Norm}$ calculated from \eqref{eq:time_7} -- which takes into account aperture roughness -- for one instance of random aperture roughness profile. These $\mathcal{P}$-field distribution calculated from \eqref{eq:time_7} for $\gamma = \SI{21}{cm}$ is plotted in Fig.~\ref{fig:sim4}. 

Comparing the $\mathcal{P}$-field integral estimate to the estimate where aperture roughness is accounted for, we observe that the $\mathcal{P}$-field integral of \eqref{eq:time3_2} provides an accurate estimate of the $\mathcal{P}$-field distribution in $\Sigma$ even for high aperture roughness values (such as our case where is of comparable dimension as $\lambda_\mathrm{P}$). For the chosen maximum roughness values of $\gamma$, the $\mathcal{P}$-field integral provides a reasonably accurate estimate of $\mathcal{P}$-field distribution despite an aperture roughness value as high as \SI{21}{cm} (70\% of the $\lambda_\mathrm{P}$). Of course, the $\mathcal{P}$-field integral provides a more accurate estimate for low roughness values and this estimate gradually degrades as $\gamma \rightarrow \lambda_\mathrm{P}$. 

To clearly show how the accuracy of the $\mathcal{P}$-field distribution estimate provided by the $\mathcal{P}$-field integral is affected by increasing roughness, we plot, in Fig.~\ref{fig:varym}, $|I_\mathrm{Tot}(x,y,z)|_\mathrm{,Norm}$ and $|I_\mathrm{Tot-F}(x,y,z)|_\mathrm{Norm}$ only along the $x$-axis in $\Sigma$ for $y=0$. For four maximum roughness values of $\gamma = \SI{3}{cm}$, $\gamma = \SI{6}{cm}$, $\gamma = \SI{9}{cm}$ and $\gamma = \SI{15}{cm}$, Fig.~\ref{fig:varym} assists us to observe that the $\mathcal{P}$-field estimate in Fig.~\ref{fig:sim1} digresses from the actual $\mathcal{P}$-field distribution when aperture roughness features become large enough to be comparable to $\lambda_\mathrm{P}$. In other words, additional $\mathcal{P}$-field specular noise, which increases with an increasing aperture roughness, is not taken into account in \eqref{eq:time3_2} which affects the Huygens-like $\mathcal{P}$-field distribution estimate in $\Sigma$ specially at large $\gamma$ values. This results in an increasing $\mathcal{P}$-field estimation error for increasing aperture roughness.

\subsection{Simulation 2: Near-Field Imaging Scenario}
\label{sec:sim2}

In Simulation 2, we repeat Simulation 1 for a near-field scenario with the location of the observation plane $\Sigma$ set to \SI{5}{m}. The dimensions of the aperture as well as the values of $\lambda_\mathrm{E}$ and $\lambda_\mathrm{P}$ are the same as in Simulation 1.  

\begin{figure}[H]
  \centering
  \subfloat[]{
  \label{fig:sim7}
  \includegraphics[width=0.3\linewidth]{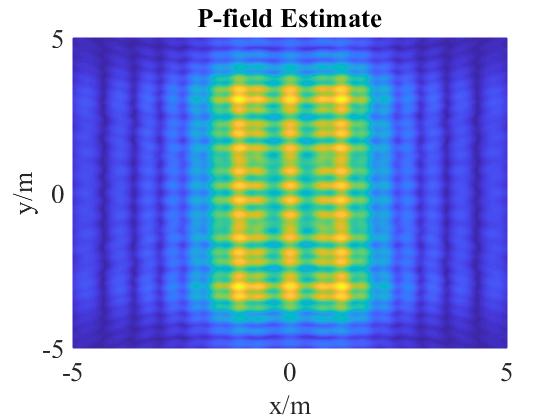}}
  \hfill
  \subfloat[]{
  \label{fig:sim8}
  \includegraphics[width=0.3\linewidth]{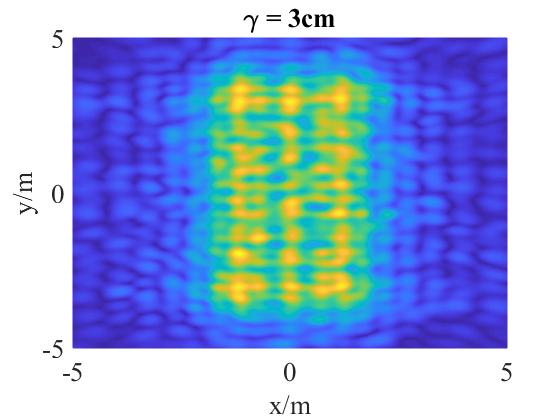}} 
  \hfill
  \subfloat[]{
  \label{fig:sim10}
  \includegraphics[width=0.3\linewidth]{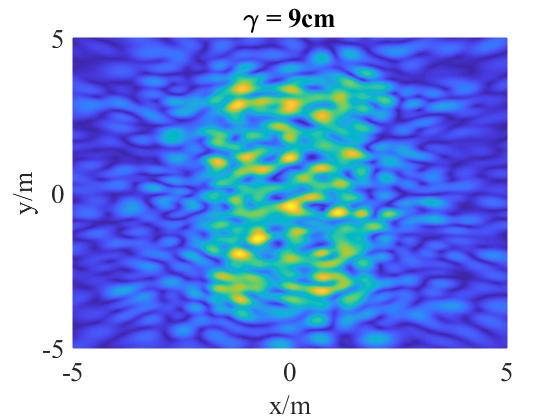}}
  \hfill
  \caption{Near-field results of Simulation 2: (a) $\mathcal{P}$-field distribution estimate $|I_\mathrm{Tot-F}(x,y,z)|_{\mathrm{Norm}}$ at $ z = \SI{5}{m}$ from the $\mathcal{P}$-field integral in \eqref{eq:time3_2}, and $\mathcal{P}$-field estimates $|I_{\mathrm{Tot}}(x,y,z)|_{\mathrm{Norm}}$ from \eqref{eq:time_7} taking into account aperture roughness of (b) $\gamma = \SI{3}{cm}$, and and (c) $\gamma = \SI{9}{cm}$.}
  \label{fig:simulations2}
\end{figure}

The $x$-$y$ plane $\mathcal{P}$-field distribution for this scenario, computed through the $\mathcal{P}$-field integral in \eqref{eq:time3_2}, is compared to the $\mathcal{P}$-field distributions calculated from \eqref{eq:time_7} for aperture roughness values of $\gamma = \SI{2}{cm}$, and $\gamma = \SI{9}{cm}$ respectively. These $\mathcal{P}$-field distributions are plotted in Fig.~\ref{fig:simulations2}. As is the case with the far-field estimates in Simulation 1, the estimate from the $\mathcal{P}$-field integral in \eqref{eq:time3_2} deteriorates with increasing roughness and because the $\mathcal{P}$-field integral assumes negligible roughness, it does not account for an increasing $\mathcal{P}$-field specular noise with an increasing aperture roughness.
%

\subsection{Simulation 3: $\mathcal{P}$-field Amplitude Estimation Error for the Ultra Near-Field Imaging Scenario}
\label{sec:sim3}
While the previous two simulations demonstrate a degradation in the $\mathcal{P}$-field distribution estimate from the $\mathcal{P}$-field integral with increasing aperture roughness, the purpose of performing Simulation 3 is to demonstrate an amplitude error incurred by the $\mathcal{P}$-field integral in the ultra near-field scenario even for negligible aperture roughness as was discussed in Section~\ref{sec:Corr_fac}. It is this scenario where the introduction of a $\mathcal{P}$-field amplitude correction factor is necessary. Through these simulations, we also show that this particular error, i.e., a dynamic scaling error with respect to the location in $\Sigma$, is independent of the phase estimation error which was investigated in Simulation 1 and Simulation 2.  

\begin{figure*}
  \centering
  \subfloat[]{
  \label{fig:Est_A}
  \includegraphics[width=0.30\linewidth]{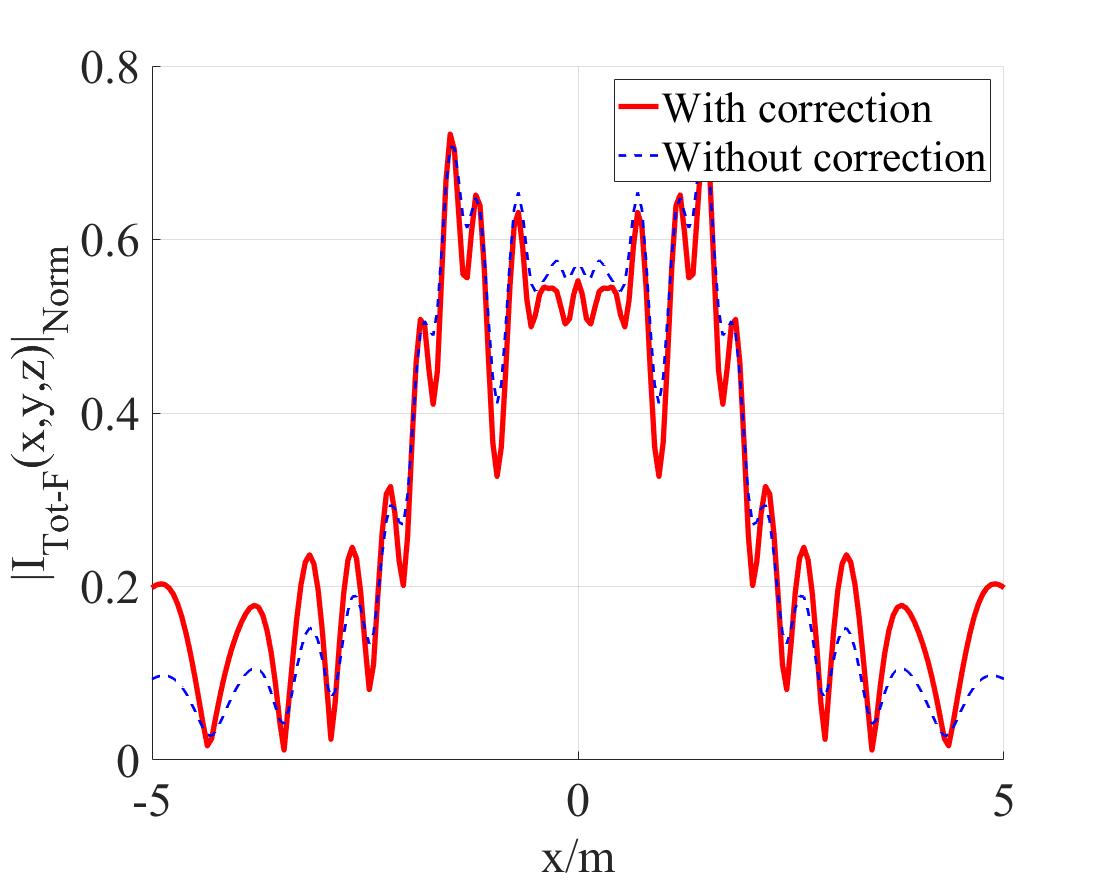}}
  \hfill
  \subfloat[]{
  \label{fig:Est_B}
  \includegraphics[width=0.30\linewidth]{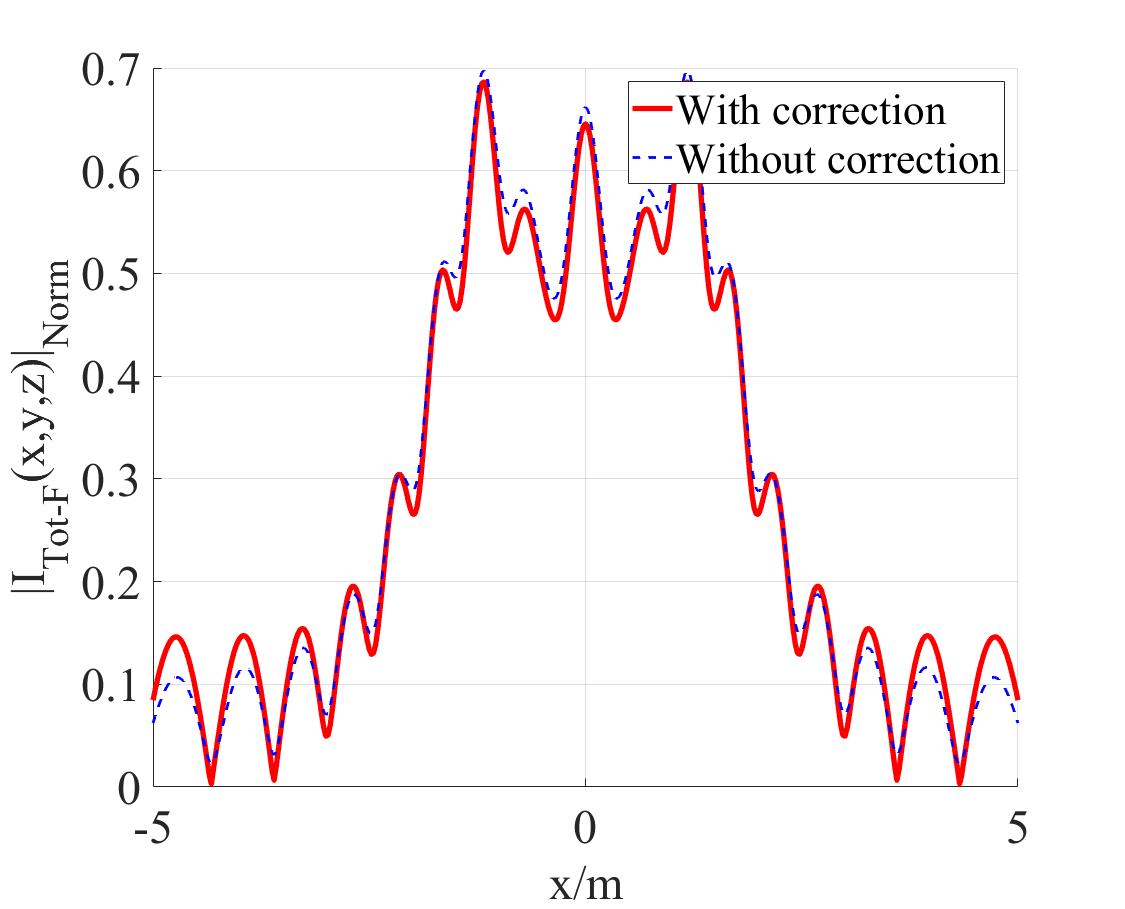}} 
  \hfill
  \subfloat[]{
  \label{fig:Est_C}
  \includegraphics[width=0.30\linewidth]{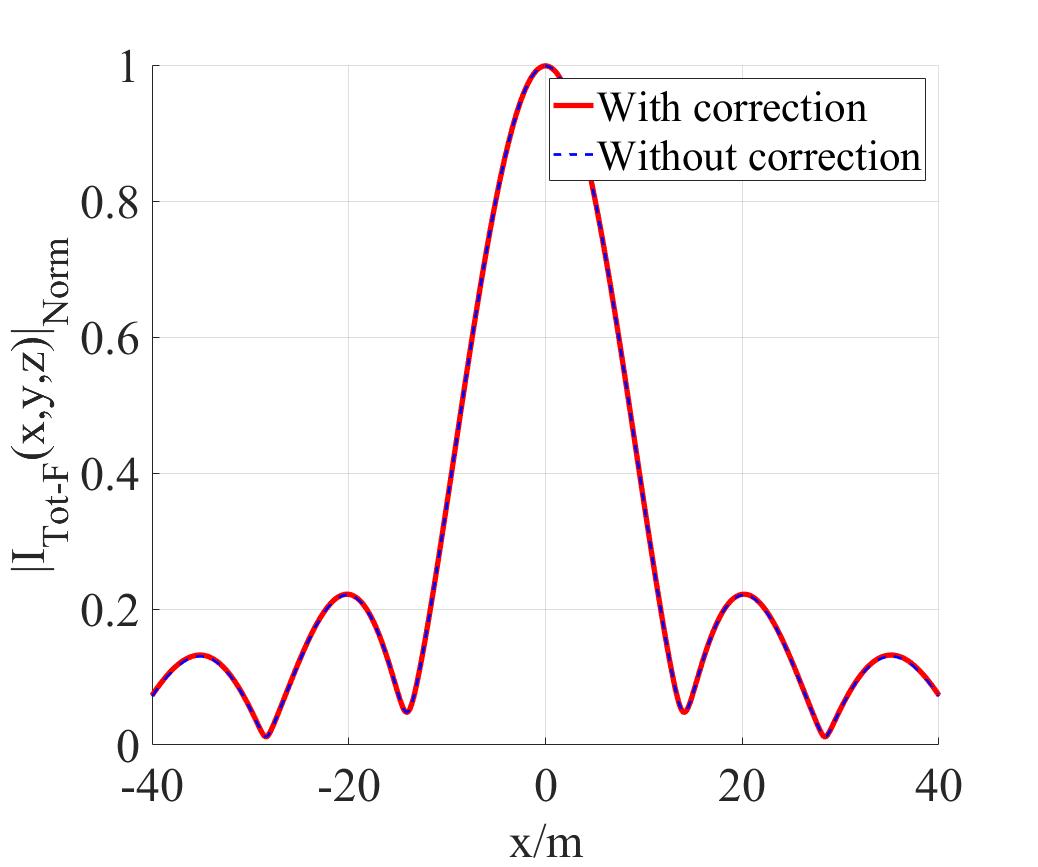}}
  \caption{Amplitude estimation error in estimates from $\mathcal{P}$-field integral with and without amplitude correction for (a) $z = \SI{2.5}{m}$, (b) $z = \SI{5}{m}$, and (c) $z = \SI{200}{m}$.}
  \label{fig:Huy_Final}
\end{figure*}

For this simulation, we first assume that the separation between $\mathcal{A}$ and $\Sigma$ is small, such that neither of the respective Fraunhofer and Fresnel conditions in \eqref{eq:Fraun_cond_a} and \eqref{eq:Fres_cond_a} apply. Next, for comparison, we also compute the amplitude error for the near-field and far-field cases of Simulations 1 and 2. To only observe the magnitude of the error in amplitude estimation, Simulation 3 considered a uniformly-illuminated aperture but with a negligible roughness, i.e., $ \gamma \approx 0$. For the ultra near-field case we set the plane separation to $z = \SI{2.5}{m}$ while the near-field and far-field distances were set again to \SI{5}{m} and \SI{200}{m} respectively. For all cases, the uncorrected $\mathcal{P}$-field distribution was estimated from \eqref{eq:time_5} and compared to the amplitude corrected estimate from \eqref{eq:PhasorWav}. These amplitude-corrected and uncorrected $\mathcal{P}$-field distributions in $\Sigma$ along the $x$-axis at $y = 0$ are plotted in Fig.~\ref{fig:Huy_Final}. 
\begin{figure*}
  \centering
  \subfloat[]{
  \label{fig:Err_A}
  \includegraphics[width=0.45\linewidth]{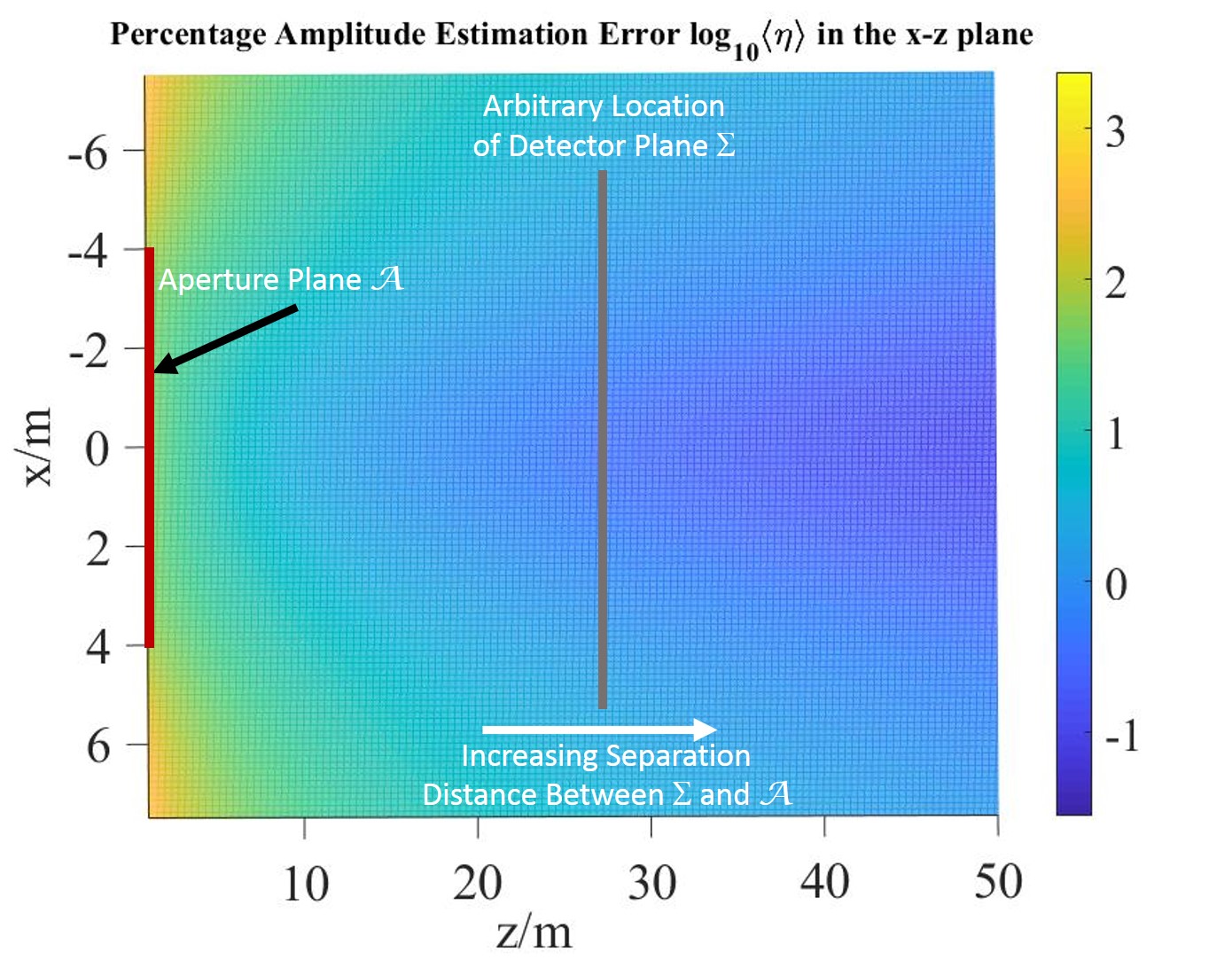}} 
  \hfill  
  \subfloat[]{
  \label{fig:Err_B}
  \includegraphics[width=0.45\linewidth]{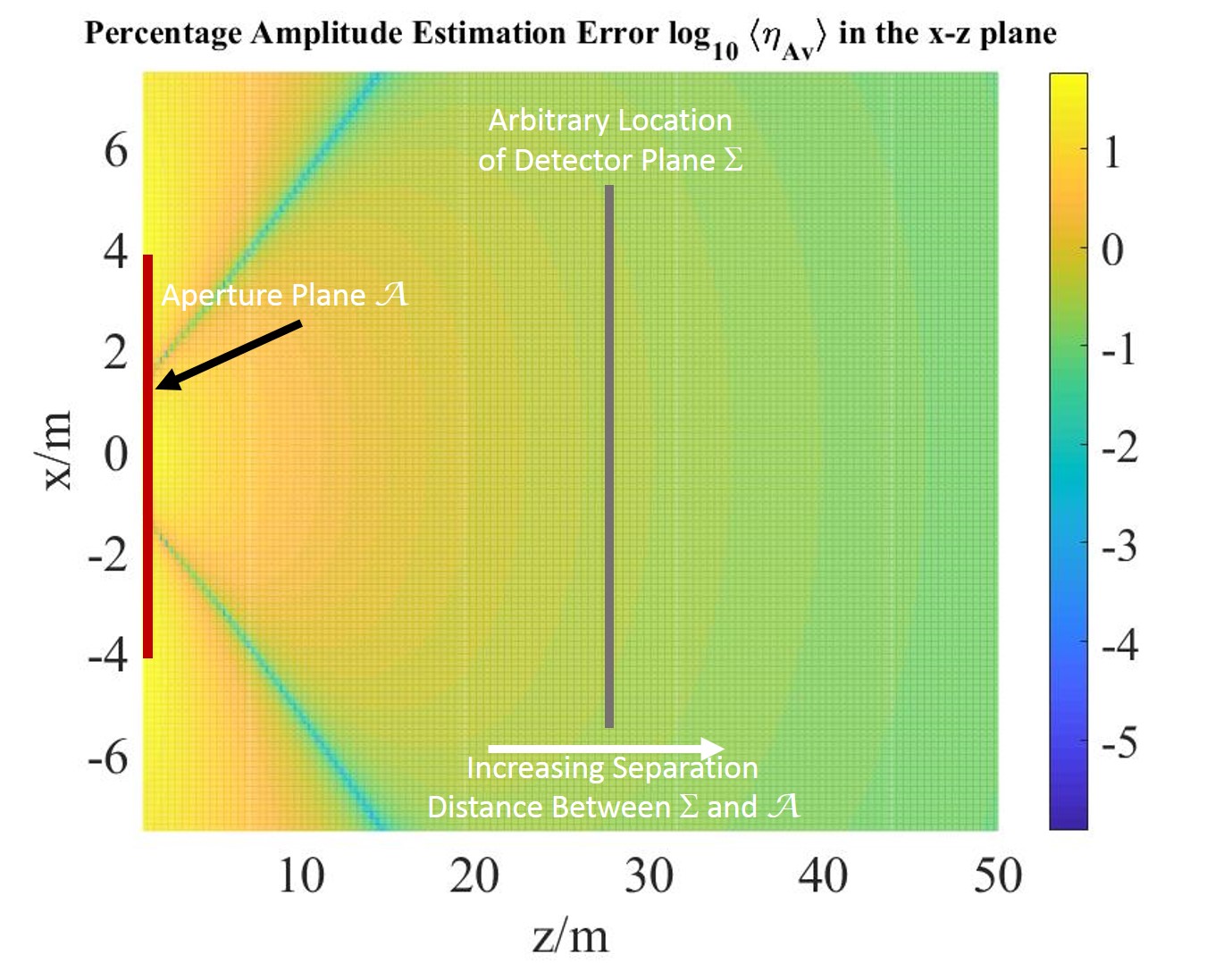}}
  \caption{A color map of (a) $\log \langle \eta  \rangle$ and (b) $\log \langle \eta_\mathrm{AV} \rangle$ for $ \SI{1}{m} \geq Z \geq \SI{50}{m}$ in the $x$-$z$ plane.} 
  \label{fig:Err_Final}
\end{figure*}

From the plots in Fig.~\ref{fig:Huy_Final}, we clearly observe that even for an ultra near-field scenario, a $\mathcal{P}$-field distribution estimate from \eqref{eq:time3_2} without the correction factor only yields an amplitude estimation error but no phase estimation error as no phase approximations were made throughout the derivation of the $\mathcal{P}$-field integral. This is evident from the locations of the minima and maxima in the $\mathcal{P}$-field distribution in Fig.~\ref{fig:Huy_Final}. 

We also observe an increasing $\mathcal{P}$-field amplitude estimation error when separation between $\mathcal{A}$ and $\Sigma$ is decreased and that it is negligible for large separation distances i.e., the Fresnel and Fraunhofer cases. This is also depicted in Fig.~\ref{fig:Err_Final} where we plot a color map of the logarithmic $\mathcal{P}$-field estimation errors $\langle \eta \rangle$ and $\langle \eta_\mathrm{AV}\rangle$ in the $x$-$z$ plane from \eqref{eq:eta1} and \eqref{eq:eta2} respectively for $\SI{50}{m} \geq z \geq \SI{1}{m}$. 

\section{Experimental Demonstration}
\label{sec:Experiments}
\begin{figure*}
  \centering
  \subfloat[]{
  \label{fig:E_int}
  \includegraphics[width=0.45\linewidth]{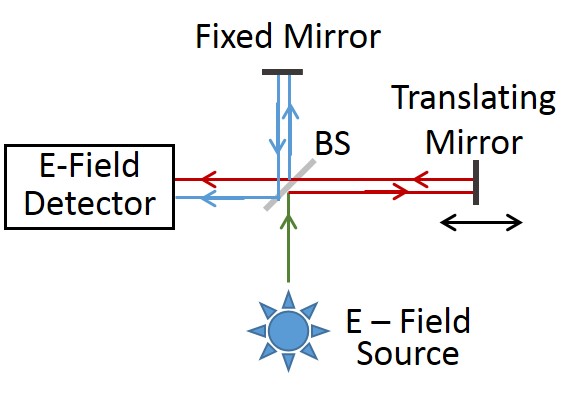}} 
  \hfill  
  \subfloat[]{
  \label{fig:P_int}
  \includegraphics[width=0.45\linewidth]{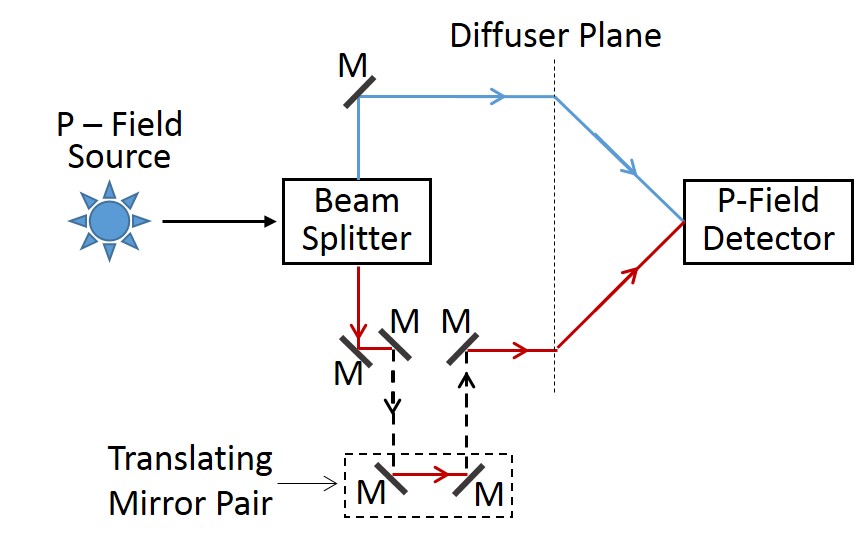}}
  \caption{Analogy between a a) classical Michelson interferometer with a movable mirror and b) the $\mathcal{P}$-field interferometer with a movable mirror pair.} 
  \label{fig:Exp_conc}
\end{figure*}

\begin{figure}
 \centering
 \includegraphics[width=0.90\linewidth]{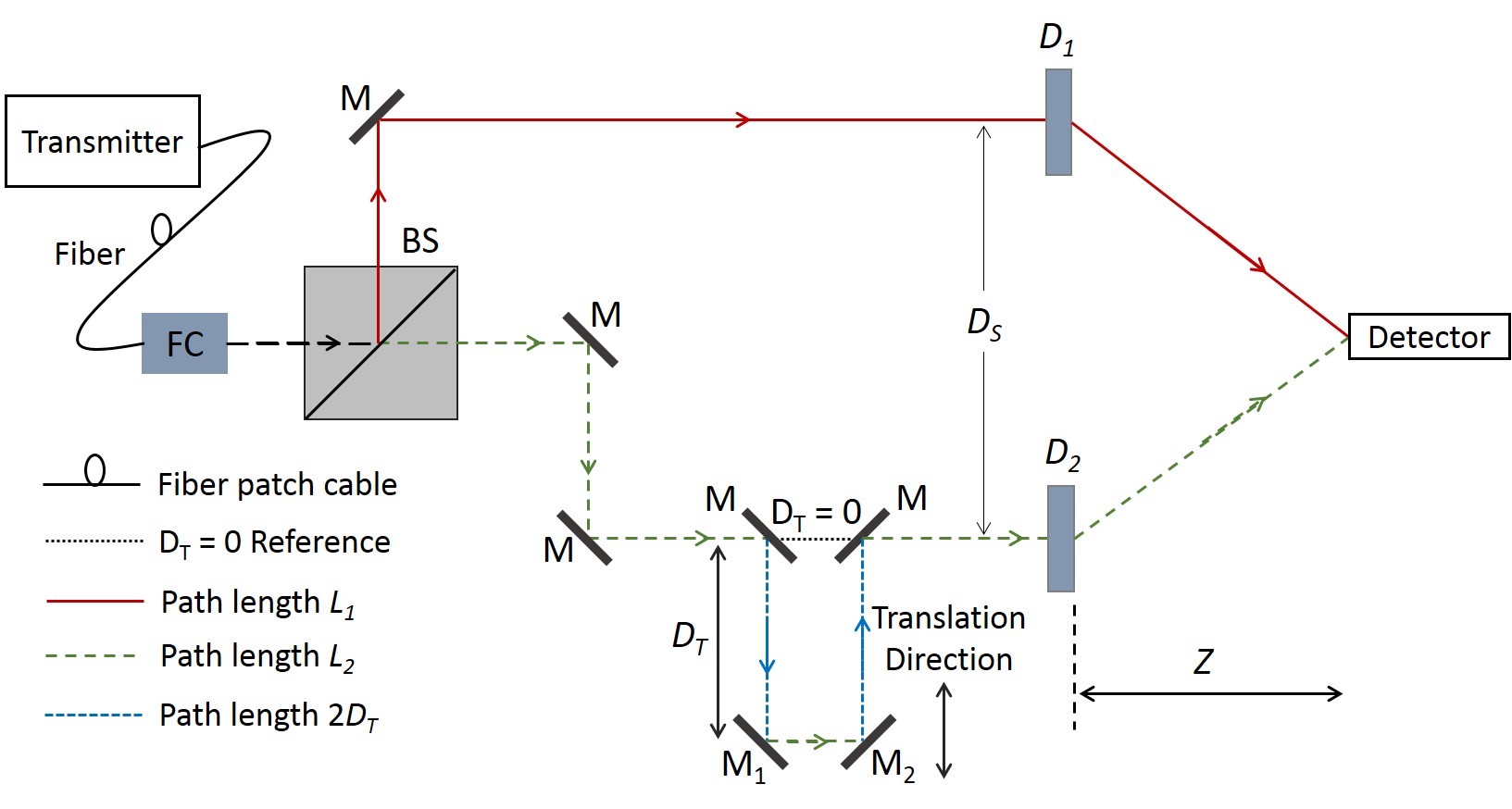}
 \caption{Experimental setup to demonstrate summation of two $\mathcal{P}$-field contributions in $\Sigma$ for different $D_\mathrm{T}$ settings.}
 \label{fig:Exp_a}
\end{figure}

We performed an experiment to verify our claim of $\mathcal{P}$-field summation behavior in the presence of rough apertures such as NLOS imaging scenarios. This, as was mentioned earlier, is analogous to the Huygens' summation of E-field wavelets to describe conventional LOS imaging. To experimentally demonstrate $\mathcal{P}$-field summation, the setup shown in Fig.~\ref{fig:Exp_a} was implemented which enacts the simplest situation of the summation of two optically incoherent $\mathcal{P}$-field contributions from a rough aperture. This experiment is analogous to an interferometry experiment in the realm of E-fields. This analogy is depicted in Fig.~\ref{fig:Exp_conc} where akin to translating a mirror and changing path length in one arm of a Michelson interferometer, we emulate the shifting of the light source with the help of a translatable mirror pair where each new location of the mirror pair introduces a unique additional phase to one beam component resulting in a different $\mathcal{P}$-field sum at a fixed location of the detector. 

A Gaussian beam from a fiber-coupled laser source exits the optical fiber cable via a Fiber Collimator (FC). A 50:50 Beam Splitter (BS) splits the propagating beam into two identical beams each with exactly half of the original beam power. We refer to these beams as Beam 1 and Beam 2. Beam 1 and Beam 2 propagate through path lengths $L_1$ and $L_2$ before incidence at identical diffusers D\textsubscript{1} and D\textsubscript{2}, separated by a distance $D_\mathrm{S}$. The path length $L_1$ serves as the reference beam path and remains unaltered throughout the course of measurements. Path length $L_2$ for Beam 2 is altered by translating a two-mirror assembly comprising mirrors M\textsubscript{1} and M\textsubscript{2} placed on a translation stage. The translation distance $D_\mathrm{T}$ of the mirror pair is measured from a reference position $D_\mathrm{T} = 0$ at which path lengths $L_1$ and $L_2$ are equal. An AC-coupled Photo-Detector (PD) is positioned at a distance \textit{Z} away from the plane of the two diffusers $\mathcal{A}$ and equidistant from D\textsubscript{1} and D\textsubscript{2}. Translation of the $[$M\textsubscript{1}, M\textsubscript{2}$]$ mirror pair resulted in altering the path length $L_2$ for Beam 2 and the consequent phase difference $\Delta\phi_\mathrm{P} = |\phi_1 - \phi_2|$ between the two $\mathcal{P}$-field contributions $\mathcal{P}_1 = |\mathcal{P}_1|e^{j\phi_1}$ and $\mathcal{P}_2 = |\mathcal{P}_2|e^{j\phi_2}$ simply expressed as a function of $D_{\mathrm{T}}$ is given by 
\begin{equation}
    \label{eq:Exp1}
    \Delta\phi_\mathrm{P} = \frac{4\pi}{\lambda_\mathrm{P}}D_\mathrm{T}.
\end{equation}
The amplitude contributions are equal, i.e., $|\mathcal{P}_2| = |\mathcal{P}_1|$ with the use of identical diffusers and a 50:50 BS. Therefore, the expected output signal $\mathcal{P}_{\mathrm{Sum}}$ which the PD outputs expressed as a function of $D_\mathrm{T}$ is given by
\begin{equation}
    \label{eq:output1}
    \mathcal{P}_{\mathrm{Sum}} =  |\mathcal{P}_1|e^{j\phi_1} + |\mathcal{P}_2|e^{j\phi_1 + \Delta\phi_\mathrm{P}}  =
    |\mathcal{P}_1|\left(1 + e^{j\frac{4\pi}{\lambda_\mathrm{P}}D_\mathrm{T}} \right).
\end{equation}
If we only detect the peak amplitude of $\mathcal{P}_{\mathrm{Sum}}$ at each $D_{\mathrm{T}}$ setting, where $|\mathcal{P}_{\mathrm{Sum}}|_{\mathrm{Peak}}$ denotes this peak amplitude of $\mathcal{P}_{\mathrm{Sum}}$, then  
\begin{equation}
    \label{eq:output2}
    |\mathcal{P}_{\mathrm{Sum}}|_{\mathrm{Peak}} = C_\mathrm{P} \bigg| |\mathcal{P}_1|e^{j\phi_1} + |\mathcal{P}_2|e^{j\phi_1 + \Delta\phi_\mathrm{P}} \bigg| =
    C_\mathrm{P} |\mathcal{P}_1|\left|1 + e^{j\frac{4\pi}{\lambda_\mathrm{P}}D_\mathrm{T}} \right|,
\end{equation}
where $C_\mathrm{P}$ is simply a coefficient of proportionality. As the maximum possible value in \eqref{eq:output2} is $2C_\mathrm{P} |\mathcal{P}_1|$, the normalized magnitude $|\mathcal{P}_{\mathrm{Norm}}|$ of the $\mathcal{P}$-field sum is given by
\begin{equation}
    \label{eq:norm_sum}
    |\mathcal{P}_{\mathrm{Norm}}| = \frac{|\mathcal{P}_{\mathrm{Sum}}|_{\mathrm{Peak}}}{2C_\mathrm{P} |\mathcal{P}_1|} = \frac{1}{2}\left|1 + e^{j\frac{4\pi}{\lambda_\mathrm{P}}D_\mathrm{T}} \right|. 
\end{equation}
As \eqref{eq:norm_sum} predicts, the two $\mathcal{P}$-field contributions from $D_1$ and $D_2$ should interfere constructively -- despite a loss in optical coherence with propagation through the rough diffusers -- when the mirror pair $[$M\textsubscript{1}, M\textsubscript{2}$]$ is located at $D_\mathrm{T} = 0$. Consequently, for any integer $Q$, we expect to detect complete destructive interference between $\mathcal{P}_1$ and $\mathcal{P}_2$ when $D_\mathrm{T} = Q\lambda_\mathrm{P} \pm \lambda_\mathrm{P}/4$. In our experiment, we observe this behavior of completely constructive and destructive interference as well as intermediate states of partial interference at different $D_\mathrm{T}$ settings.

For the actual experiments, we used a fiber-coupled laser source with a wavelength of $\lambda_\mathrm{E} = \SI{520}{nm}$ modulated by a $\mathcal{P}$-field sinusoidal signal of \SI{1}{GHz} which corresponds to a free-space $\mathcal{P}$-field wavelength of roughly \SI{30}{cm}. The diffusers were separated by a distance of $D_\mathrm{S}$ of \SI{34.5}{cm} and the distance $Z$ between the aperture plane $\mathcal{A}$ (plane of the diffusers) and the detection plane $\Sigma$ was set to \SI{36}{cm}. Two identical diffusers from the Newport 20DKIT-C3 light-shaping diffusers kit were used which provided a Gaussian irradiance distribution at the detector plane. A $\mathcal{P}$-field detector to measure $|\mathcal{P}_{\mathrm{Norm}}|$ was realized by connecting the output of an AC-coupled Menlo System APD210 photo-detector to an Agilent CXA N9000A RF spectrum analyzer. The spectrum analyzer output $|\mathcal{P}_{\mathrm{Sum}}|_{\mathrm{Peak}}$ was normalized in post-processing to obtain $|\mathcal{P}_{\mathrm{Norm}}|$.

\begin{figure}[pt]
 \centering
 \includegraphics[width=0.90\linewidth]{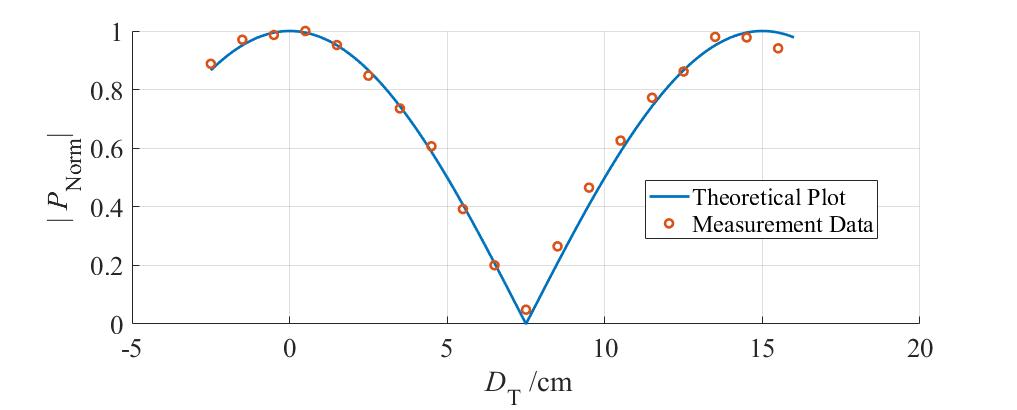}
 \caption{Comparison between theoretical predictions and experimental measurements of $|\mathcal{P}_{\mathrm{Norm}}|$ for different $D_\mathrm{T}$ settings.}
 \label{fig:Exp_b}
 \includegraphics[width=0.9\linewidth]{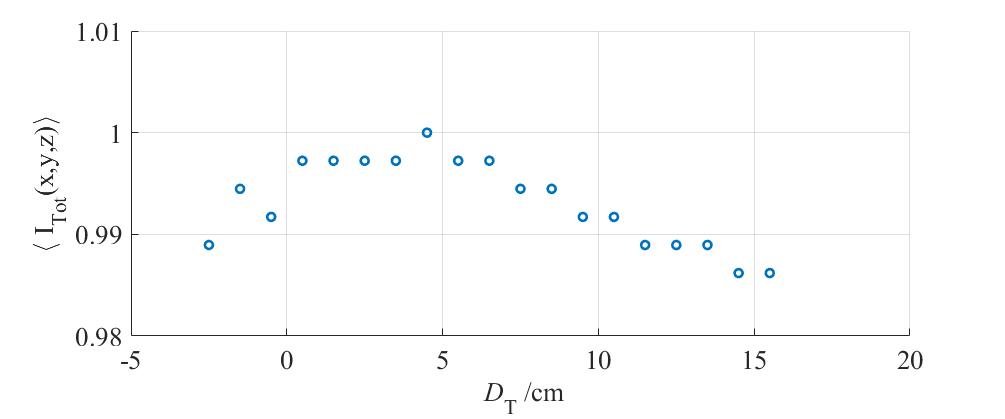}
 \caption{Measurements of $\langle I(x,y,z) \rangle$ for different $D_\mathrm{T}$ settings.}
 \label{fig:Exp_c}
\end{figure}

The mirror pair $[$M\textsubscript{1}, M\textsubscript{2}$]$ was translated and $|\mathcal{P}_{\mathrm{Norm}}|$ was measured for different unique values of $D_\mathrm{T}$. In Fig.~\ref{fig:Exp_b}, we plot simultaneously the measurement data-point values of $|\mathcal{P}_{\mathrm{Norm}}|$ as well as a plot of the theoretically expected behavior of $|\mathcal{P}_{\mathrm{Norm}}|$ as a function of $D_\mathrm{T}$ calculated from \eqref{eq:norm_sum}. Comparing the theoretically predicted and experimentally measured values of $|\mathcal{P}_{\mathrm{Norm}}|$ at different $D_\mathrm{T}$ settings, we clearly observe a maximum $\mathcal{P}$-field summation and cancellation at the expected $D_\mathrm{T}$ settings as well as a very strong agreement for other $D_\mathrm{T}$ settings which yield varying levels of partial $\mathcal{P}$-field interference. 

Using a slow Thorlabs SM05PD1A power meter/PD assembly with an integration time that is far greater than the time period $\mathcal{P}$-field signal, we also recorded simultaneous measurements of the total optical irradiance at the location of the fast PD. Through these measurements, we show that a change in the $\mathcal{P}$-field values at different $D_\mathrm{T}$ settings is independent of the total number of photons present at the detection location which we expected to remain almost constant for any $D_\mathrm{T}$. We provide a plot of this average optical irradiance in Fig.~\ref{fig:Exp_c}. As was expected, the normalized average irradiance $\langle I_\mathrm{Tot}(x,y,z) \rangle$ remains almost constant for all $D_\mathrm{T}$ settings.

%

\section{$\mathcal{P}$-field Imaging Approach to NLOS Imaging}
\label{sec:NLOS_App_Main}

\subsection{Comparison of NLOS Imaging with a $\mathcal{P}$-field Virtual Camera to LOS Imaging with a Conventional Camera}
\label{sec:NLOS_App2}

\begin{figure*}
  \centering
  \subfloat[]{
  \label{fig:conventional_a}
  \includegraphics[width=0.45\linewidth]{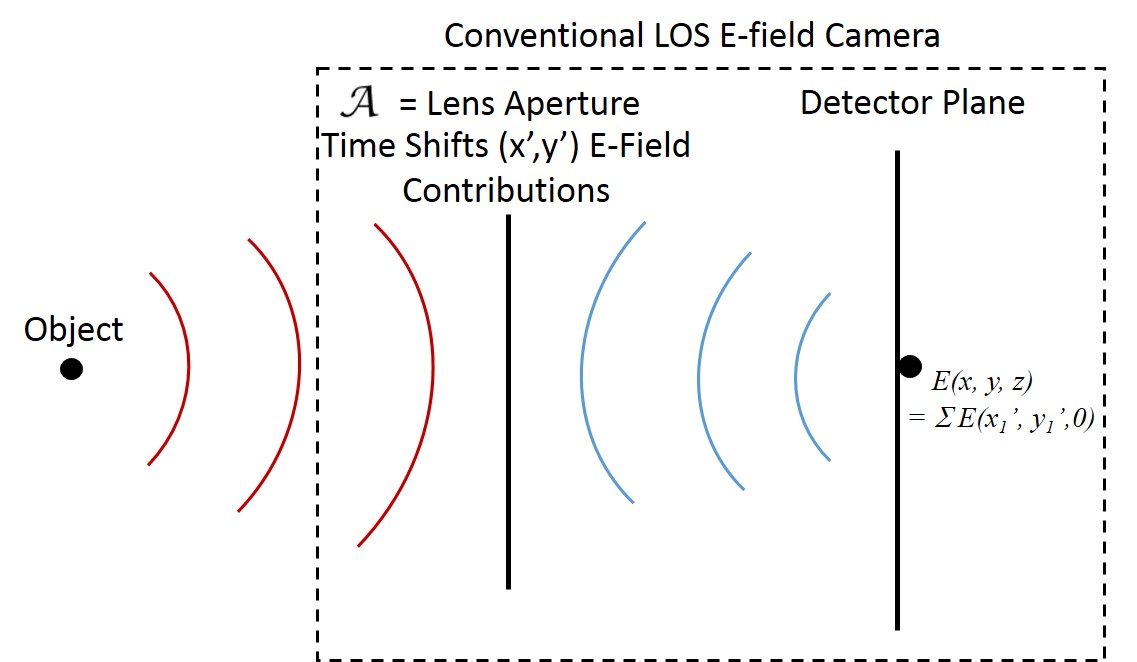}} 
  \hfill  
  \subfloat[]{
  \label{fig:conventional_b}
  \includegraphics[width=0.45\linewidth]{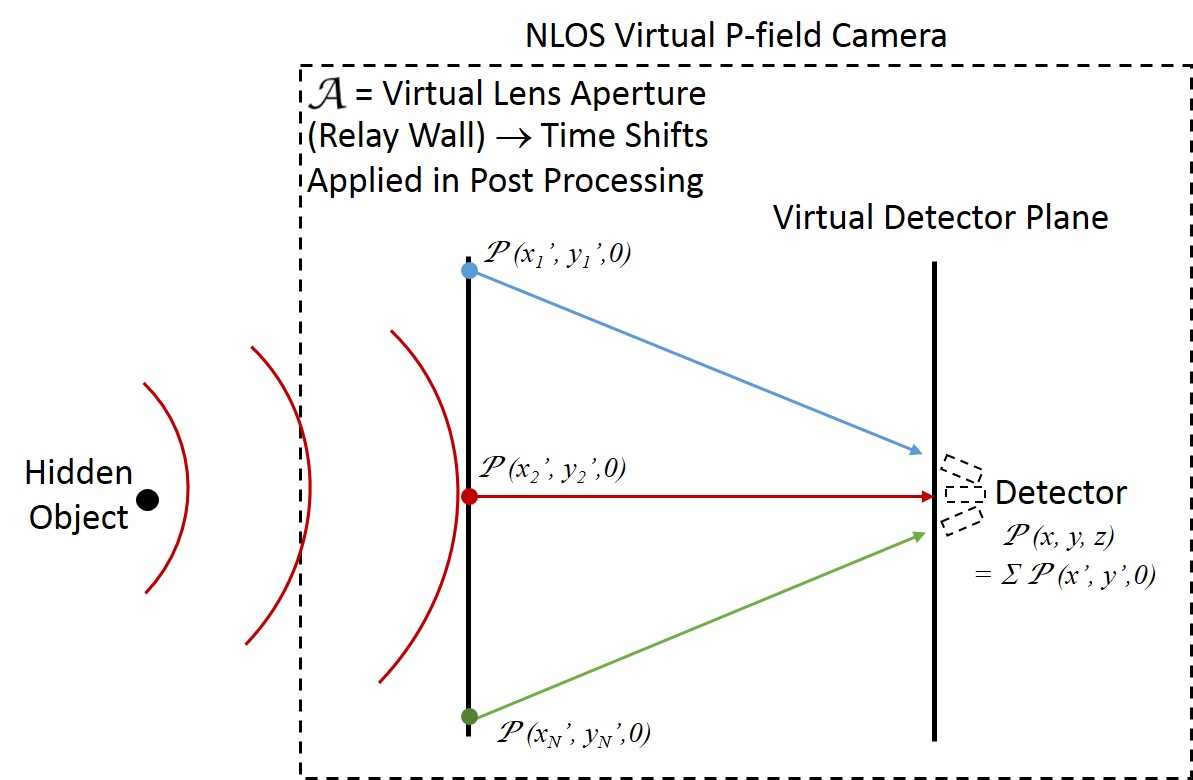}}
  \caption{Comparison between a) conventional LOS E-field imaging and b) $\mathcal{P}$-field based NLOS imaging.} 
  \label{fig:conventional}
\end{figure*}

Fundamental E-field imaging principles defined by the Huygens' integral in \eqref{eq:E_wav3} completely describe conventional LOS imaging, as is depicted in Fig.~\ref{fig:conventional_a}, where we consider an imaging system comprising a monochromatic EM source, a camera and a point-like target (which we consider our object under investigation). The monochromatic wave emitted by the source interacts with the object, and is focused at the detector/camera plane by an E-field imaging lens.

In a virtual $\mathcal{P}$-field camera approach for NLOS imaging, the relay wall acts as the aperture of a holographic $\mathcal{P}$-field projection and detection system. After detection of the $\mathcal{P}$-field at all points on the aperture, any conceivable imaging system can be realized through digital post processing. A simple imaging lens, for example, applies a position dependent phase delay to the signal followed by a summation of the fields on the camera pixels and a time integral over the absolute value to reconstruct a 2D image of a scene (Fig.~\ref{fig:nlos}). With the application of correct time shifts to the received signal, the relay wall can be treated as a virtual $\mathcal{P}$-field lens, which forms a $\mathcal{P}$-field image of the NLOS at the virtual sensor located behind the relay wall looking directly at the hidden scene. This is depicted in Fig.~\ref{fig:conventional_b}.

\subsection{Limitations of the Phasor-Field model}
In the derivation of the Phasor-Field Rayleigh Sommerfeld propagator we made two important assumptions:
\begin{itemize}
    \item Light is added incoherently such that the intensities of two overlapping light beams add.
    \item Light propagates from any point in the the source plane on a spherical wavefront in all directions.
\end{itemize}

Another way to say this is that the light used is sufficiently incoherent that the phase of the electric field is random on scales shorter than the Phasor-Field wavelength. To assure this, we require that the source surface is diffuse which creates spatial incoherence. If the source surface is specular or partially specular, i.e. if it does not randomize the carrier phase, the Phasor-Field model no longer completely describes light transport. This situation arises when the phasor field wave propagates around an occluding object surrounded by free space. To simulate propagation in this case one would have to propagate the field to the plane of the occluding object and from there to the destination surface behind the occluder. Since the plane of the occluding object contains free space, Phasor-Field light propagation does not correctly predict the field behind the object. A similar situation arises when imaging a NLOS object that is a specular reflector. In these cases, a different model or extension of the Phasor-Field formalism is required. An extension that can model occlusions has been introduced by Dove et. al.~\cite{dove2019paraxial}.

It is interesting to note that, while specular reflectors and occlusions in the scene affect the validity of the Phasor-Field  model, they do not generally affect reconstructions done using the Phasor-Field formalism which look the same regardless of the surface specularies~\cite{liu2018virtual}. However, in the presence of occlusions and specularities it is in certain cases possible to perform NLOS reconstructions with higher quality than what is permitted using Phasor-Field methods.

\section{Conclusion}
\label{sec:conclusions}

In this paper, we introduced the concept of irradiance \emph{phasor fields}, a complex scalar quantity with an amplitude and a phase term. We show that $\mathcal{P}$-fields provide an ideal representation for imaging through apertures which exhibit roughness which is in the order of the optical (E-field) wavelength but significantly less than the $\mathcal{P}$-field wavelength. Provided that the optical spatial coherence of the E-field is less than the $\mathcal{P}$-field wavelength and its temporal coherence is less than the detector integration time, the proposed $\mathcal{P}$-field approach provides a means to model any NLOS imaging system -- such as imagers that image around corners -- as an LOS imaging system. The significant advantage with the $\mathcal{P}$-fields-based approach is the inherent ability to use well-known techniques in LOS imaging to model any NLOS system by considering partially reflective and scattering rough surfaces as $\mathcal{P}$-field apertures and using existing knowledge of light transport in LOS imaging to model any NLOS system from there onward. We back our claims with in-depth near-field and far-field simulation results for uniformly illuminated rectangular and square apertures as well as results from carefully designed experiments. 

\newpage

\section*{Appendix: $\mathcal{P}$-field Imaging Approach}
\label{sec:Appendix_A}
\setcounter{equation}{0}
\renewcommand{\theequation}{A.{\arabic{equation}}}
\subsection*{Irradiance Propagation between Two Spatial Locations}
\label{sec:Poyn}

The Poynting vector $\Vec{S}$, which describes energy propagation of any electro-magnetic (EM) wave, is stated as a cross-product of the wave electric field (E-field) vector $\Vec{E}$ and the magnetic field vector $\Vec{H}$ as
\begin{equation}
\label{eq:Poyn1}
    \Vec{S} = \Vec{E}\times\Vec{H}.
\end{equation}
Moreover, $\Vec{H}$ can be expressed in terms of the wave vector $\Vec{K}$, $\Vec{E}$, the E-field frequency $\omega$ and the medium permeability $\mu$ as  
\begin{equation}
    \label{eq:Def1}
    \Vec{H} = \frac{\Vec{K}\times\Vec{E}}{\omega\mu},
\end{equation}
where $\Vec{K}$ can be expressed either in terms of the E-field wavelength $\lambda_\mathrm{E}$ or likewise $\omega$ as
\begin{equation}
    \label{eq:Def_K} 
    \Vec{K} = \frac{2\pi}{\lambda_\mathrm{E}}\Hat{K} = \frac{\omega}{c}\Hat{K}.
\end{equation}
In \eqref{eq:Def_K}, $\Hat{K}$ signifies the unit vector $\Vec{K}/|\Vec{K}|$. Substituting \eqref{eq:Def1} into \eqref{eq:Poyn1}, and knowing that $\Vec{E}\cdot\Vec{K} = 0$ for an isotropic medium, $\Vec{S}$ is given by 
\begin{equation}
    \label{eq:poyn2}
    \Vec{S}=\frac{1}{\omega\mu}\left[\Vec{K}\left(\Vec{E}\cdot\Vec{E}\right)\right]=\frac{|\Vec{K}|}{\omega\mu}\left(\Vec{E}\cdot\Vec{E}\right)\Hat{K} = \frac{\left(\Vec{E}\cdot\Vec{E}\right)}{\zeta} \Hat{K}
\end{equation}
using the triple cross-product identity
\begin{equation}
    \label{eq:cross}
    \Vec{a}\times\left(\Vec{b}\times\Vec{c} \right) = \Vec{b}\left(\Vec{a}\cdot\Vec{c}\right)-\Vec{c}\left(\Vec{a}\cdot\Vec{b}\right).
\end{equation}
In \eqref{eq:poyn2} the quantity $\zeta$ is the impedance of the propagation medium given by
\begin{equation}
    \label{eq:def2}
    \zeta=\frac{\omega\mu}{|\Vec{K}|}.
\end{equation}
For time-harmonic E-fields, a similar approach enables the derivation of an expression for the time-averaged Poynting vector $\langle\Vec{S}\rangle$ for optical irradiance \cite{silver1984microwave} where
\begin{equation}
    \label{eq:Poyn_spec}
    \Vec{\langle S \rangle} = \frac{1}{T} \int_0^T \Vec{S}dt = \frac{1}{2} \frac{|\Vec{K}|}{\omega\mu} \operatorname{Re}[\Vec{E}_\mathrm{S}\cdot \Vec{E_\mathrm{S}^{\ast}}] \Hat{K} = \frac{1}{2} \frac{\Vec{E_\mathrm{S}}\cdot \Vec{E_\mathrm{S}^{\ast}}}{\zeta} \Hat{K}. 
\end{equation}
In \eqref{eq:Poyn_spec}, $\Vec{E}_\mathrm{S}$ is simply $\Vec{E}$ expressed in the phasor notation. Note that although the expressions in \eqref{eq:poyn2} and \eqref{eq:Poyn_spec} are similar, yet they are distinctly different quantities i.e., $\Vec{S}$ is a time-varying Poynting vector and $\langle\Vec{S}\rangle$ is a time-averaged Poynting vector which is usually what a typical photo-detector detects.

Next, we use this knowledge to a) rigorously derive the irradiance contribution from a location $(x',y',0)$ within a plane $\mathcal{A}$ -- which we refer to as the plane of a rough aperture (see Section C of this appendix for definition of 'roughness' that we use for our considerations) -- to a location $(x,y,z)$ in the $\Sigma$ plane which we refer to as the detection plane or the image plane as is shown in Fig.~\ref{fig:huygens}, and b) derive the cumulative irradiance contribution of all locations within $\mathcal{A}$ to location $(x,y,z)$ in $\Sigma$. Let 
\begin{equation}
    \label{eq:E-Field2}
    \Vec{E}_{\mathrm{Inc}} (x',y',0) = E_0(x',y',0)\cos{\left(\omega t\right)}\Hat{e'}
\end{equation}
be the incident E-field at $(x',y',0)$ in $\mathcal{A}$ where $\Hat{e'}$ denotes the E-field polarization unit vector. Then the transmitted E-field contribution from the same location $(x',y',0)$ in $\mathcal{A}$ is given by
\begin{equation}
    \label{eq:E-Field3}
    \Vec{E}(x',y',0) = t_0(x',y',0) E_0(x',y',0)\cos{\left(\omega t + \Delta \phi_{K}(x',y',0) \right)}\Hat{e},
\end{equation}
where $t_0(x',y',0)$ is the magnitude and  $\Delta\phi_{K}(x',y',0)$ denotes the random phase contribution to the transmission coefficient due to aperture roughness at location $(x',y',0)$ of the E-field transmission coefficient $\tau (x',y',0) = t_0(x',y',0)e^{j\Delta\phi_\mathrm{K}(x',y',0)}$.  The unit vector $\Hat{e}$ denotes the polarization of the transmitted E-field which can be different from the initial polarization $\Hat{e'}$ of $\Vec{E}_{\mathrm{Inc}} (x',y',0)$. Moreover, owing to its time-harmonic nature, $\Vec{E}(x',y',0)$ can be expressed as a phasor quantity (where subscript S denotes phasor-equivalent notation for the E-field contribution from $(x',y',0) \in \mathcal{A}$)
\begin{equation}
    \label{eq:def_phasor2}
    \Vec{E}_\mathrm{S}(x',y',0) = t_0(x',y',0) E_0(x',y',0)e^{j\Delta\phi_{K}(x',y',0)} \Hat{e}.
\end{equation}
The E-field contribution from $(x',y',0) \in \mathcal{A}$ to $(x,y,z) \in \Sigma$ is expressed as a spherical E-field wavelet $\mathcal{E}$ with a $1/|r|$ amplitude drop-off and a $\phi_{K} = K|r|$ phase accumulated due to propagation. $\mathcal{E}$ is given by
\begin{equation}
    \label{eq:E-Field}
    \mathcal{E} = \Vec{E}(x,y,z) = \frac{t_0(x',y',0) E_0(x',y',0)}{|r|}\cos{\left(\omega t + K|r| + \Delta\phi_{K}(x',y',0) \right)}\Hat{e},
\end{equation}
where $|r|=\sqrt{\left(x-x'\right)^2+\left(y-y'\right)^2+z^2}$ is the distance between locations $(x',y',0)$ and $(x,y,z)$, and $K = |\Vec{K}| = \omega/c$ is the E-field wave number. As was the case in \eqref{eq:def_phasor2}, $\Vec{E}(x,y,z)$ can also be expressed in the phasor notation as $\Vec{E}_\mathrm{S}(x,y,z)$ where
\begin{equation}
    \label{eq:def_phasor1}
    \Vec{E}_\mathrm{S}(x,y,z) = \frac{1}{|r|} \underbrace{t_0(x',y',0) E_0(x',y',0)e^{j \left[ \Delta\phi_{K}(x',y',0)\right] } e^{jK|r|} \Hat{e}}_{\Vec{E}_M(x,y,z)} = \frac{1}{|r|}\Vec{E}_\mathrm{M}(x,y,z).
\end{equation}
In \eqref{eq:def_phasor1}, $\Vec{E}_\mathrm{M}(x,y,z)$ is the phasor representation of the E-field $\Vec{E}_\mathrm{S}(x,y,z)$ contributed from $(x',y',0) \in \mathcal{A}$ without the $1/|r|$ amplitude drop-off. In other words, we simply define a new quantity $\Vec{E}_\mathrm{M}(x,y,z)$ (still in the phasor notation) as the E-field $\Vec{E}_\mathrm{S}(x',y',0)$ from location $(x',y',0) \in \mathcal{A}$ with the propagation phase added but without the amplitude drop-off i.e.
\begin{equation}
    \label{eq:E_M}
    \Vec{E}_\mathrm{M}(x,y,z) = \Vec{E}_\mathrm{S}(x',y',0) e^{j \left( K|r| \right)}.
\end{equation}
It follows from \eqref{eq:Poyn_spec} and \eqref{eq:E_M}, that the corresponding time-averaged optical irradiance contribution $I(x',y',0)$ emanating from aperture location $(x',y',0)$ can be expressed both in terms of $\Vec{E}_\mathrm{S}(x',y',0)$ or $\Vec{E}_\mathrm{M}(x,y,z)$ as 
\begin{equation}
    \label{eq:irradiance1}
    I(x',y',0) = \frac{\Vec{E}_\mathrm{S}(x',y',0)\cdot\Vec{E}_\mathrm{S}^{\ast}(x',y',0)}{2\zeta}  = \frac{\Vec{E}_\mathrm{M}(x,y,z) \cdot \Vec{E}_\mathrm{M}^{\ast}(x,y,z)}{2\zeta} = \frac{|E_0(x',y',0)|^2}{2\zeta}.
\end{equation}
Similarly, the corresponding time-averaged optical irradiance contribution $I(x,y,z)$ at $(x,y,z) \in \Sigma$ from $(x',y',0) \in \mathcal{A}$ is expressed as 
\begin{equation}
    \label{eq:irradiance2}
    I(x,y,z) = \frac{\Vec{E}_\mathrm{S}(x,y,z)\cdot\Vec{E}_\mathrm{S}^{\ast}(x,y,z)}{2\zeta} = \frac{ \Vec{E}_\mathrm{M}(x,y,z) \cdot \Vec{E}_\mathrm{M}^{\ast}(x,y,z)}{2\zeta|r|^2} = \frac{I(x',y',0)}{|r|^2}\,.
\end{equation}
Having defined this basic E-field framework for this manuscript, we now proceed to derive the $\mathcal{P}$-field integral in the following section. 

\subsection*{$\mathcal{P}$-field Propagation between Aperture and Detection Planes}
\label{sec:newframework}

Now let us consider the case when optical irradiance is directly amplitude-modulated. A direct amplitude modulation of optical irradiance is expressed as the multiplication of the time-varying Poynting vector $\Vec{S}$ by a non-negative scalar modulating function $P(t)$ of frequency $\Omega$ and an amplitude of $P_0$ where
\begin{equation}
    \label{eq:P_Field_Def2}
    P\left(t\right) = P_0 \left[1+\cos{\left(\Omega t \right)}\right].
\end{equation}
After propagation from $(x',y',0)$ to $(x,y,z)$, the modulation envelope $P(t)$ is phase-shifted by $\beta |r|$ due to propagation and $\Delta\phi_{\beta}(x',y',0)$ due to aperture roughness at $(x',y',0)$. This phase-shifted envelope at $(x,y,z)$ is expressed as 
\begin{equation}
    \label{eq:P_Field_Def}
    P\left(r,t\right) = P_0\left[1+\cos{\left(\Omega t + \beta |r| + \Delta\phi_{\beta}(x',y',0) \right)}\right].
\end{equation}
In \eqref{eq:P_Field_Def}, $\beta = 2\pi/\lambda_\mathrm{P}$ is the wave number of $P(r,t)$ expressed in terms of its wavelength $\lambda_\mathrm{P}$. We demonstrate later that this modulating function $P(r,t)$ scaled by the time-averaged unmodulated optical irradiance contribution is a 'Huygens-like' phasor field ($\mathcal{P}$-field) contribution from a diffuse aperture considered rough as per the definitions in Section~\ref{sec:roughness}. We can also write \eqref{eq:P_Field_Def} as 
\begin{equation}
    \label{eq:P_Field_def2}
    P\left(r,t\right)=2P_0\cos^2\left(\frac{\Omega t}{2} + \frac{\beta |r|}{2} +\frac{\Delta\phi_{\beta}(x',y',0)}{2} \right),    
\end{equation}
which we can also express as
\begin{equation}
    \label{eq:P_notation}
    P(r,t) = \left( \operatorname{Re}\left[P_\mathrm{M}e^{\frac{j\Omega t}{2}}\right]\right) \left( \operatorname{Re}\left[P_\mathrm{M}e^{\frac{j\Omega t}{2}}\right]\right).
\end{equation}
The spatially-dependent phasor representation of $P(r,t)$ in \eqref{eq:P_notation} is denoted by $P_M$ which is stated as
\begin{equation}
    \label{eq:P_notation2}
    P_\mathrm{M} = \sqrt{2P_0}e^{j \left( \beta |r|/2 + \Delta\phi_{\beta}(x',y',0)/2 \right) }.
\end{equation}
In the following steps - in order to avoid long expressions - we denote $\Vec{E}_\mathrm{M}(x,y,z)$ as simply $\Vec{E}_\mathrm{M}$. From the definitions of $\Vec{E}_\mathrm{M}$ and $P_\mathrm{M}$ in \eqref{eq:E_M} and \eqref{eq:P_notation2} respectively, the time-averaged Poynting vector $\langle S(x,y,z) \rangle$ signifying the amplitude modulated irradiance contribution at $(x,y,z)$ from $(x',y',0)$ is expressed as
\begin{multline}
    \label{eq:Poyn4}
     \langle\Vec{S}(x,y,z)\rangle = \frac{\mathcal{L}}{|r|^2\zeta} \bigg[ \frac{1}{T}\int_0^T \left( \operatorname{Re}\left[\Vec{E}_\mathrm{M} e^{j\omega t}\right]\, \operatorname{Re} \left[P_\mathrm{M} e^{j\Omega t /2}\right]\right)\cdot  \left(\operatorname{Re}\left[\Vec{E}_\mathrm{M} e^{j\omega t}\right]\, \operatorname{Re} \left[P_\mathrm{M} e^{j\Omega t /2}\right]\right) dt \bigg] \Hat{K},  
\end{multline}
where $\mathcal{L}$ -- a loss factor -- is introduced to account for irradiance loss due to scattering/absorption at the aperture and in propagation from $(x',y',0)$ to $(x,y,z)$ and $T$ is the detector integration time. Consequently, $\langle\Vec{S}(x,y,z)\rangle$ can be simply expressed as
\begin{multline}
    \label{eq:Poyn5}
    \langle\Vec{S}(x,y,z)\rangle = \frac{\mathcal{L}}{|r|^2\zeta} \bigg[ \frac{1}{T}\int_0^T \bigg(\frac{\Vec{E}_\mathrm{M} e^{j\omega t} + \Vec{E}_\mathrm{M}^{\ast} e^{-j\omega t}}{2}\bigg)\bigg(\frac{P_\mathrm{M} e^{j\Omega t/2} + P_\mathrm{M}^{\ast} e^{-j\Omega t/2}}{2}\bigg) \cdot \\
    \bigg(\frac{\Vec{E}_\mathrm{M} e^{j\omega t} + \Vec{E}_\mathrm{M}^{\ast} e^{-j\omega t}}{2}\bigg)\bigg(\frac{P_\mathrm{M} e^{j\Omega t/2} + P_\mathrm{M}^{\ast} e^{-j\Omega t/2}}{2} \bigg) dt \bigg]\Hat{K}.
\end{multline}
It is to be noted that a photo-detector is only able to measure changes to the time-averaged irradiance magnitude $I(x,y,z)$ of the Poynting vector. Therefore, from hereon, we omit the unit vector $\Hat{K}$ signifying the direction of electro-magnetic energy flow and denote $|\langle \Vec{S}(x,y,z) \rangle| = I(x,y,z)$. For the case when the average aperture roughness is larger than the optical wavelength $\lambda_\mathrm{E}$ and where the spatial coherence of the optical carrier is consequently reduced to less than the $\mathcal{P}$-field wavelength $\lambda_\mathrm{P}$, the total irradiance contribution $I_\mathrm{Tot}(x,y,z)$ from all aperture locations to a single location $(x,y,z)$ in $\Sigma$ can be expressed as the sum of all independent irradiance contributions from each location within $\mathcal{A}$ i.e.,
\begin{equation}
    \label{eq:incoh_sum}
    I_{\mathrm{Tot}}(x,y,z) \propto \int_{\mathcal{A}}\left( \frac{1}{T} \int_0^T  \frac{I(x',y',0)}{|r|^2} dt\right) dx'dy'.
\end{equation}
Defining $\Omega' \coloneqq \Omega/2$ in \eqref{eq:Poyn5} and expanding the expression in \eqref{eq:Poyn5} using the incoherent irradiance summation of \eqref{eq:incoh_sum}, $I_{\mathrm{Tot}}(x,y,z)$ can be expressed as
\begin{multline}
      \label{eq:Poyn7}
     I_{\mathrm{Tot}}(x,y,z) \approx \left( \frac{\mathcal{L}K_\mathrm{P}}{16\zeta}\right) \int_{\mathcal{A}} \frac{1}{|r|^2} \bigg( \frac{1}{T}\int_0^T \bigg[\underbrace{ \Vec{E}_\mathrm{M} \cdot \Vec{E}_\mathrm{M} e^{2j\omega t} P_\mathrm{M} P_\mathrm{M} e^{2j\Omega' t}}_{\mathrm{Term 1}} + \\ + \underbrace{ \Vec{E}_\mathrm{M}^{\ast} \cdot \Vec{E}_\mathrm{M}^{\ast} e^{-2j\omega t} P_\mathrm{M}^{\ast} P_\mathrm{M}^{\ast} e^{-2j\Omega' t}}_{\mathrm{Term 2}}  
    + \underbrace{ \Vec{E}_\mathrm{M} \cdot \Vec{E}_\mathrm{M} e^{2j\omega t} P_\mathrm{M} P_\mathrm{M}^{\ast}}_{\mathrm{Term 3}} + \underbrace{ \Vec{E}_\mathrm{M}^{\ast} \cdot \Vec{E}_\mathrm{M}^{\ast} e^{-2j\omega t} P_\mathrm{M}^{\ast} P_\mathrm{M}}_{\mathrm{Term 4}} + \\
    + \underbrace{ \Vec{E}_\mathrm{M} \cdot \Vec{E_M} e^{2j\omega t} P_\mathrm{M}^{\ast} P_\mathrm{M}}_{\mathrm{Term 5}} + \underbrace{ \Vec{E}_\mathrm{M}^{\ast} \cdot \Vec{E}_\mathrm{M}^{\ast} e^{-2j\omega t} P_\mathrm{M} P_\mathrm{M}^{\ast}}_{\mathrm{Term 6}} + \\ 
    + \underbrace{ \Vec{E}_\mathrm{M} \cdot \Vec{E}_\mathrm{M} e^{2j\omega t} P_\mathrm{M}^{\ast} P_\mathrm{M}^{\ast} e^{-2j\Omega' t}}_{\mathrm{Term 7}}
    + \underbrace{ \Vec{E}_\mathrm{M}^{\ast} \cdot \Vec{E}_\mathrm{M}^{\ast} e^{-2j\omega t} P_\mathrm{M} P_\mathrm{M} e^{2j\Omega' t}}_{\mathrm{Term 8}} \bigg] dt \bigg) dx'dy' \\
     + \\
     \left( \frac{\mathcal{L}K_\mathrm{P}}{16\zeta} \right) \int_{\mathcal{A}} \frac{1}{|r|^2} \bigg( \frac{1}{T} \int_0^T \bigg[ \underbrace{ \Vec{E}_\mathrm{M} \cdot \Vec{E}_\mathrm{M}^{\ast} P_\mathrm{M} P_\mathrm{M} e^{2j\Omega' t}}_{\mathrm{Term 9}} + \underbrace{ \Vec{E}_\mathrm{M}^{\ast} \cdot \Vec{E}_\mathrm{M} P_\mathrm{M}^{\ast} P_\mathrm{M}^{\ast} e^{-2j\Omega' t}}_{\mathrm{Term 10}} \\ +  \underbrace{\Vec{E}_\mathrm{M} \cdot \Vec{E}_\mathrm{M}^{\ast} P_\mathrm{M} P_\mathrm{M}^{\ast}}_{\mathrm{Term 11}}  
     + \underbrace{ \Vec{E}_\mathrm{M}^{\ast} \cdot \Vec{E}_\mathrm{M} P_\mathrm{M}^{\ast} P_\mathrm{M}}_{\mathrm{Term 12}} \\ 
     + \underbrace{ \Vec{E}_\mathrm{M} \cdot \Vec{E}_\mathrm{M}^{\ast} P_\mathrm{M}^{\ast} P_\mathrm{M}^{\ast} e^{-2j\Omega' t}}_{\mathrm{Term 13}} +  \underbrace{ \Vec{E}_\mathrm{M}^{\ast} \cdot \Vec{E}_\mathrm{M} P_\mathrm{M} P_\mathrm{M} e^{2j\Omega' t}}_{\mathrm{Term 14}} + \\
     + \underbrace{ \Vec{E}_\mathrm{M} \cdot \Vec{E}_\mathrm{M}^{\ast} P_\mathrm{M}^{\ast} P_\mathrm{M}}_{\mathrm{Term 15}} + \underbrace{ \Vec{E}_\mathrm{M}^{\ast} \cdot \Vec{E}_\mathrm{M} P_\mathrm{M} P_\mathrm{M}^{\ast}}_{\mathrm{Term 16}}
      \bigg] dt \bigg)dx'dy'.
\end{multline}
In \eqref{eq:Poyn7}, $K_\mathrm{P}$ is a coefficient of proportionality. For a detector with an integration time \textit{T} where
\begin{equation}
    \label{eq:int_time}
    \frac{2\pi}{\omega} \ll T \ll \frac{2\pi}{\Omega}, 
\end{equation}
the detector integration time is much longer than the time period of the extremely high frequency  $2\omega$ and allows the time-average of terms 1--8 in \eqref{eq:Poyn7} to be approximated as zero.
Recognizing that terms 10, 12, 14, 16 are the respective complex conjugates of terms 9, 11, 13, 15, we can state
\begin{multline}
    \label{eq:Av_Poyn2}
       I_{\mathrm{Tot}}(x,y,z) \approx \left( \frac{K_\mathrm{P} \mathcal{L}}{16\zeta}\right) \int_{\mathcal{A}} \bigg[ \frac{1}{T} \int_t^{t+T} \frac{1}{|r|^2} \bigg( 2\operatorname{Re} \left[ \Vec{E}_\mathrm{M} \cdot \Vec{E}_\mathrm{M}^{\ast} P_\mathrm{M} P_\mathrm{M} e^{2j\Omega' \tau }\right] + \\ 
       2\operatorname{Re} \left[ \Vec{E}_\mathrm{M} \cdot \Vec{E}_\mathrm{M}^{\ast} P_\mathrm{M} P_\mathrm{M}^{\ast} \right]
    + 2\operatorname{Re} \left[ \Vec{E}_\mathrm{M} \cdot \Vec{E}_\mathrm{M}^{\ast} P_\mathrm{M}^{\ast} P_\mathrm{M} \right] +
    \\ + 2\operatorname{Re} \left[ \Vec{E}_\mathrm{M} \cdot \Vec{E}_\mathrm{M}^{\ast} P_\mathrm{M}^{\ast} P_\mathrm{M}^{\ast} e^{-2j\Omega' \tau }\right] \bigg) d\tau\bigg]dx'dy'.  
\end{multline}
Moreover, $I_{\mathrm{Tot}}(x,y,z)$ is expressed as $I_{\mathrm{Tot}}(x,y,z,t)$ because the slowly changing irradiance envelope still results in a time-dependence of different irradiance measurements over a fixed duration of the detector integration time window. Substituting \eqref{eq:irradiance1} in \eqref{eq:Av_Poyn2}, we obtain 
\begin{multline}
    \label{eq:Av_Poyn_temp1}
   I_{\mathrm{Tot}}(x,y,z,t) \approx \frac{K_\mathrm{P} \mathcal{L}}{4} \int_{\mathcal{A}}  \bigg[\frac{1}{T}\int_t^{t+T} \frac{I(x',y',0)}{|r|^2} \bigg( \operatorname{Re} \left[ P_\mathrm{M} P_\mathrm{M} e^{2j\Omega' \tau }\right] + \operatorname{Re} \left[ P_\mathrm{M} P_\mathrm{M}^{\ast} \right] + \\
    + \operatorname{Re} \left[ P_\mathrm{M}^{\ast} P_\mathrm{M} \right] +  \operatorname{Re} \left[ P_\mathrm{M}^{\ast} P_\mathrm{M}^{\ast} e^{-2j\Omega' \tau }\right]  \bigg) d\tau\bigg] dx'dy'.
\end{multline}
Substituting $P_\mathrm{M}$ from \eqref{eq:P_notation2}, we can express \eqref{eq:Av_Poyn_temp1} as
\begin{multline}
    \label{eq:Av_Poyn6}
    I_{\mathrm{Tot}}(x,y,z,t) \approx \frac{K_\mathrm{P} \mathcal{L}}{2} \int_{\mathcal{A}} \bigg[ \frac{1}{T}\int_t^{t+T} \frac{1}{|r|^2} \bigg( P_0I(x',y',0) \cos{(\Omega \tau + \beta |r| + \Delta\phi_{\beta}(x',y',0))}  + \\
    + 2P_0I(x',y',0) +  P_0I(x',y',0) \cos{(-\Omega \tau - \beta |r| - \Delta\phi_{\beta}(x',y',0))}  \bigg) d\tau\bigg] dx'dy'.
\end{multline}
If the $\mathcal{P}$-field amplitude contribution $\mathcal{P}_{0,\Omega}(x',y',z)$ from any aperture location $(x',y',0)$ is defined as $\mathcal{P}_{0,\Omega}(x',y',0) = I(x',y',0)P_0$,
\begin{multline}
    \label{eq:Av_Poyn7}
    I_{\mathrm{Tot}}(x,y,z,t) \approx K_\mathrm{P} \mathcal{L} \int_{\mathcal{A}} \bigg[ \frac{1}{T}\int_t^{t+T} \frac{1}{|r|^2}\bigg(\mathcal{P}_{0,\Omega}(x',y',0) [1+\cos{(\Omega \tau + \beta |r| +\Delta\phi_{\beta}(x',y',0))}] \bigg) d\tau\bigg] dx'dy'.
\end{multline}
When an AC-coupled photo-detector is used for detection, the DC components of all $\mathcal{P}$-field contributions filter out which allows us to detect signed $\mathcal{P}$-field contributions expressed as  
\begin{equation}
    \label{eq:Av_Poyn10}
    I_{\mathrm{Tot}}(x,y,z,t) \approx K_\mathrm{P} \mathcal{L} \int_{\mathcal{A}} \bigg[ \frac{1}{T}\int_t^{t+T} \frac{1}{|r|^2}\bigg(\mathcal{P}_{0,\Omega}(x',y',0) \cos{[\Omega \tau + \beta |r| + \Delta\phi_{\beta}(x',y',0)]} \bigg) d\tau\bigg] dx'dy'.
\end{equation}
Considering the fact that $T\ll 2\pi/\Omega$, the cosine with frequency $\Omega$ approximately does not change within the interval $T$, so we can write
\begin{equation}
    \label{eq:time3}
    I_{\mathrm{Tot}}(x,y,z,t) \approx K_\mathrm{P} \mathcal{L} \int_{\mathcal{A}}  \frac{1}{|r|^2}\bigg(\mathcal{P}_{0,\Omega}(x',y',0) \cos{[\Omega t + \beta |r| + \Delta\phi_{\beta}(x',y',0)]} \bigg)  dx'dy'.
\end{equation}
Moreover, attributing to the time-harmonic nature of $\mathcal{P}$-fields, $I_{Tot}(x,y,z,t)$ is expressed as a sum of $\mathcal{P}$-field wavelet contributions ${\mathcal{P}(r)}$ as
\begin{equation}
    \label{eq:time_4}
    I_{\mathrm{Tot}}(x,y,z) \approx K_\mathrm{P} \mathcal{L} \int_{\mathcal{A}} \frac{1}{|r|} \underbrace{\mathcal{P}_{0,\Omega}(x',y',0)\frac{e^{j\beta |r|}}{|r|}}_{\mathcal{P}(r)} e^{j\Delta\phi_{\beta}(x',y',0)} dx'dy',
\end{equation}
where $\mathcal{P}(r)$ is the phasor-equivalent representation of each $\mathcal{P}$-field wavelet contribution $\mathcal{P}(r,t)$ given by
\begin{equation}
    \label{eq:p_def2}
    \mathcal{P}(r,t)=\frac{\mathcal{P}_{0,\Omega}(x',y',0)}{|r|} \cos{(\Omega t + \beta |r|)}.
\end{equation}
As long as the E-field spatial coherence is less than the $\mathcal{P}$-field wavelength and the E-field temporal coherence is less than the detector integration time, this sum of $\mathcal{P}$-field contributions in \eqref{eq:time_4} completely describes light transport through rough apertures (such as apertures used in NLOS scenarios) in a Huygens-like formulation (shown in \eqref{eq:E_wav3} as a sum of E-field wavelet contributions) which describes conventional LOS imaging.

Suppose we define a detection scheme - which we shall refer to as a $\mathcal{P}$-field detector - which simply detects the scalar peak amplitude $|I_{\mathrm{Tot}}(x,y,z)|$ of the $\mathcal{P}$-field envelope $I_{\mathrm{Tot}}(x,y,z)$ at every location $(x,y,z)$ in $\Sigma$ where
\begin{equation}
    \label{eq:time_7}
    |I_{\mathrm{Tot}}(x,y,z)| = \bigg|K_\mathrm{P} \mathcal{L} \int_{\mathcal{A}} \frac{1}{|r|}\mathcal{P}(r) e^{j\Delta\phi_{\beta}(x',y',0)} dx'dy'\bigg|.
\end{equation}
Moreover, when either the near-field or the far-field approximation
\begin{equation}
 \label{eq:Fraun_cond_a}
 z \gg \frac{\pi}{\lambda_\mathrm{P}} \, (x^2 + y^2 + x'^2 + y'^2)
\end{equation}
\begin{equation}
 \label{eq:Fres_cond_a}
 z^3 \gg  \frac{\pi}{4\lambda_\mathrm{P}} \left[(x - x')^2 + (y - y')^2 \right]_{\mathrm{Max}}^2,
\end{equation}
hold, then \eqref{eq:time_7} can be expressed as
\begin{equation}
    \label{eq:time_5_2}
    |I_\mathrm{Tot}(x,y,z)| = \bigg|\frac{K_\mathrm{P}\mathcal{L}}{z} \int_{\mathcal{A}} \left[ \mathcal{P}_{0,\Omega}(x',y',0)\frac{e^{j\beta |r|}}{|r|}\right] e^{j\Delta\phi_{\beta}(x',y',0)} dx'dy'\bigg|.
\end{equation}
The subscript 'Max' in \eqref{eq:Fres_cond_a} refers to the maximum possible value of the sum of coordinates inside of the brackets. On the other hand, when neither of the near-field or far-field approximations are applicable, then \eqref{eq:time_5_2}, with the inclusion of a $\mathcal{P}$-field amplitude correction factor $C(x,y,z,|r|)$ as shown in Sec.\ref{sec:Corr_fac}, can be expressed as
\begin{equation}
    \label{eq:time_6_2}
    |I_\mathrm{Tot}(x,y,z)| = \bigg|\frac{K_\mathrm{P}\mathcal{L}C(x,y,z,|r|)}{|r(x,y,z)|_{\mathrm{Av}}} \int_{\mathcal{A}} \left[ \mathcal{P}_{0,\Omega}(x',y',0)\frac{e^{j\beta |r|}}{|r|}\right]e^{j\Delta\phi_{\beta}(x',y',0)} dx'dy'\bigg|.
\end{equation}
Furthermore, if the aperture roughness is considered negligible for the case when $\lambda_\mathrm{E} \ll \lambda_\mathrm{P}$ $\implies \Delta\phi_{\beta} \approx 0 \, \text{ } \forall (x',y',0) \in \mathcal{A}$, we can express \eqref{eq:time_7} as
\begin{equation}
    \label{eq:time3_2}
    |I_{\mathrm{Tot-F}}(x,y,z)| = \bigg|K_\mathrm{P} \mathcal{L} \int_{\mathcal{A}} \frac{1}{|r|}\mathcal{P}(r) dx'dy'\bigg|.
\end{equation}
Moreover, we can express \eqref{eq:time3_2} analogously to \eqref{eq:time_5_2} and \eqref{eq:time_6_2} for the far-field/near-field and the ultra near-field cases respectively as
\begin{equation}
    \label{eq:time_3_3}
       |I_\mathrm{Tot-F}(x,y,z)| = \bigg|\frac{K_\mathrm{P} \mathcal{L}}{z} \int_{\mathcal{A}} \mathcal{P}(r) dx'dy'\bigg|,
\end{equation}
and
\begin{equation}
    \label{eq:time_7_2}
    |I_\mathrm{Tot-F}(x,y,z)| = \bigg|\frac{K_\mathrm{P}\mathcal{L}C(x,y,z,|r|)}{|r(x,y,z)|_{\mathrm{Av}}} \int_{\mathcal{A}} \mathcal{P}(r) dx'dy'\bigg|.
\end{equation}
We refer to \eqref{eq:time_3_3} as the $\mathcal{P}$-field integral for the near-field and the far-field cases and \eqref{eq:time_7_2} as the $\mathcal{P}$-field integral for the ultra near-field imaging case where the additional subscript F denotes a quasi-flat relay wall with enough roughness to randomize the E-field phase but insufficient to incur any amount of significant random phase change to the $\mathcal{P}$-field contributions.

\subsection*{Impact of Aperture Roughness on $\mathcal{P}$-field Phase}
\label{sec:roughness}

For the propagation of modulated optical irradiance from an aperture plane $\mathcal{A}$ to the detection plane $\Sigma$, we consider the relative effect of the aperture roughness on the E-field and $\mathcal{P}$-field phases. For the visible spectrum, a frosted glass or a lens with a ground glass side are examples of partially-transmissive rough apertures. On the other hand, a painted wall is one example of a partially-reflective rough aperture. Now let us assume that for a rough aperture (refer to Fig.\ref{fig:huygens}), the E-field transmission function $\tau(x',y',0)$ is given by
\begin{equation}
 \label{eq:T_rand}
 \tau(x', y',0) = t_0(x', y',0) \, e^{j \Delta \phi_R (x',y',0)},
\end{equation}
where $t_0(x',y',0)$ is the location-dependent E-field amplitude transmissivity of the aperture and $\Delta \phi_R (x',y',0)$ is a random phase variable denoting a random E-field phase contribution from any location $(x',y',0)$. In this paper, we consider an aperture 'rough' if any random aperture phase contribution from an arbitrary location $(x',y',0) \in \mathcal{A}$ results in a corresponding random phase change $\Delta\phi_{K} (x',y',0)$ to the E-field contribution at $(x',y',0)$ yet the resulting $\mathcal{P}$-field phase change $\Delta\phi_{\beta}(x',y',0)$ is negligible, i.e., $\Delta\phi_{\beta} (x',y',0) \ll 2\pi$. Mathematically speaking, without loss of generality, we assume that the surface has a minimum roughness of 0 and a maximum roughness of $\gamma$ (having the unit of length) with $\lambda_\mathrm{E} \ll \gamma \ll \lambda_\mathrm{P}$. We therefore know that 
\begin{equation}
 \label{eq:delta_phaseE}
 \Delta\phi_{K}(x',y',0)\leq\frac{2\pi}{\lambda_{E}}\gamma
\end{equation}
and
\begin{equation}
 \label{eq:delta_phaseP}
 \Delta\phi_\beta(x',y',0)\leq\frac{2\pi}{\lambda_{P}}\gamma\,.
\end{equation}
This implies that
\begin{equation}
 \label{eq:delta_phaseP_1}
 \Delta\phi_\beta(x',y',0) = \frac{\lambda_\mathrm{E}}{\lambda_\mathrm{P}} \Delta\phi_K(x',y',0).
\end{equation}
From \eqref{eq:delta_phaseP_1}, given $\lambda_\mathrm{E} \ll \gamma \ll \lambda_\mathrm{P}$, it can be inferred that the E-field phase shift at any aperture location is randomly distributed in the interval $[0,2\pi]$ while the $\mathcal{P}$-field phase shift 
\begin{equation}
 \label{eq:delta_phaseP_final}
 \Delta\phi_\beta(x',y',0)\approx 0\,.
\end{equation}
This means that an aperture surface with roughness greater than the E-field wavelength but lesser than the $\mathcal{P}$-field wavelength results in randomizing the E-field phase, while the $\mathcal{P}$-field phase remains unaltered.


\section*{Acknowledgments}
The authors thank Jeremy Teichman for valuable discussions and suggestions.


\section*{Funding}
This work was funded by DARPA through the DARPA REVEAL project (HR0011-16-C-0025), the NASA Innovative Advanced Concepts (NIAC) program (NNX15AQ29G), and the Office of Naval Research (ONR, N00014-15-1-2652).


\bibliography{bibliography}

\end{document}